# Cyber Risk at the Edge: Current and future trends on Cyber Risk Analytics and Artificial Intelligence in the Industrial Internet of Things and Industry 4.0 Supply Chains

Petar Radanliev[1], David De Roure[1]; Kevin Page[1]; Jason R.C. Nurse[2]; Rafael Mantilla Montalvo[3]; Omar Santos[3]; La'Treall Maddox[3]; Pete Burnap[4]

[1]Oxford e-Research Centre, Department of Engineering Sciences, University of Oxford, UK; petar.radanliev@oerc.ox.ac.uk; [2]School of Computing, University of Kent, UK; [3]Cisco Research Centre, Research Triangle Park, USA; [4]School of Computer Science and Informatics, Cardiff University, Wales, UK.

**Abstract:** Digital technologies have changed the way supply chain operations are structured. In this article, we conduct systematic syntheses of literature on the impact of new technologies on supply chains and the related cyber risks. A taxonomic/cladistic approach is used for the evaluations of progress in the area of supply chain integration in the Industrial Internet of Things and Industry 4.0, with a specific focus on the mitigation of cyber risks. An analytical framework is presented, based on a critical assessment with respect to issues related to new types of cyber risk and the integration of supply chains with new technologies. This paper identifies a dynamic and self-adapting supply chain system supported with Artificial Intelligence and Machine Learning (AI/ML) and real-time intelligence for predictive cyber risk analytics. The system is integrated into a cognition engine that enables predictive cyber risk analytics with real-time intelligence from IoT networks at the edge. This enhances capacities and assist in the creation of a comprehensive understanding of the opportunities and threats that arise when edge computing nodes are deployed, and when AI/ML technologies are migrated to the periphery of IoT networks.

Pre-print – before proofread by journal print production team.
Reference:

Radanliev, Petar, David De Roure, Kevin Page, Jason R.C. Nurse, Rafael Mantilla Montalvo, Omar Santos, La'Treall Maddox, and Pete Burnap. "Cyber Risk at the Edge: Current and Future Trends on Cyber Risk Analytics and Artificial Intelligence in the Industrial Internet of Things and Industry 4.0 Supply Chains." *Cybersecurity, Springer Nature*, 2020. https://doi.org/10.1186/s42400-020-00052-8.

*Keywords: Industry 4.0; Supply Chain Design; Transformational Design Roadmap; IIoT Supply Chain Model; Decision Support for Information Management, Artificial Intelligence and Machine Learning (AI/ML), dynamic self-adapting system, cognition engine, predictive cyber risk analytics.*

# 1 Introduction

There are many businesses opportunities in networking supply chains within the new digital economy (Bauer, Hämmerle, Schlund, & Vocke, 2015). Smart manufacturing is set to create large resource savings (G. Anderson, 2016), and enable economies of scale (Brettel, Fischer, Bendig, Weber, & Wolff, 2016). The new paradigm of Industry 4.0 (I4.0) will enable organisations to meet individual customer requirements and create value opportunities (B. Lee, Cooper, Hands, & Coulton, 2019b), increase resource productivity, and provide flexibility in businesses processes (Hussain, 2017). To allow for this however, it requires integration of the Industrial-Internet-of Things (IIoT) theories, control of physical systems, and modelling interaction between humans and Cyber Physical Systems (CPS) (Marwedel & Engel, 2016). Business and supply chain models need to embrace the opportunities from I4.0 (Jazdi, 2014; Wahlster et al., 2013), for enhancing and automating their businesses process decomposition and real-world visibility. Real-time enabled CPS and IIoT platforms should represent the foundation for I4.0 businesses and respective supply chain models (Marwedel & Engel, 2016). The idea of I4.0 was introduced with the development of IIoT and CPS (Ashton, 2011; Gershenfeld, 1999). The IIoT and CPS have sought to integrate the real and virtual worlds together (Tan, Goddard, & Pérez, 2008), promoting automation with real-time enabled platforms (Ringert, Rumpe, & Wortmann, 2015).

Although there is a consensus on the value from embracing the I4.0 (Shafiq, Sanin, Szczerbicki, & Toro, 2015), the impact of cyber risk remains to be determined (Okutan, Werner, Yang, & McConky, 2018). There has been some advancements however with automation of vulnerability discovery (Y. Wang et al., 2019), and ensuring data confidentiality and secure deletion (Zhang, Jia, Chang, & Chen, 2018). The IIoT and Supply Chain Management in I4.0 need to prepare for high-grade digitisation of processes, smart manufacturing, and inter-company connectivity (Müller, Buliga, & Voigt, 2018). This requires understanding of the relationship between technological entrepreneurship and socio-economic changes (L. Li, 2017).

A key novelty of this study is the process of using IoT design principles, presented as a step-by-step transformational roadmap. Technology road-mapping of information and communication technologies (ICT) is present in literature (Bloem da Silveira Junior, Vasconcellos, Vasconcellos Guedes, Guedes, & Costa, 2018). The findings from this study are building upon previous work on understanding the I4.0 trends for key smart manufacturing technologies (Lu & Weng, 2018), and contribute for policy development.

The article builds upon existing studies on attack synthesis and towards predictive cyber defence (Okutan & Yang, 2019) and graph-based visual analytics for cyber threat intelligence (Böhm, Menges, & Pernul, 2018), but distinguishes between ICT and IIoT. This is considered as fundamental distinction for narrowing the research efforts on understanding how modern IIoT technological concepts can be integrated in I4.0 supply chains.

We review how artificial intelligence and IoT introduce new challenges to privacy, security and resilience of connected supply chain environments. This study builds upon the FAIR institute (FAIR, 2020) methodology by redefining the FAIR institute definition on 'explicit' risk management. The research focuses on how AI methods can be used to increase or decrease the precision and scale of attacks, by automating aspects such as intelligence gathering, target selection, and attack execution. The IoT devices built into digital supply chains greatly increase the amount of data captured. This could result in data leaks and significant privacy risks. While this topic is widely debated, less research has been conducted on how AI techniques and IoT devices could strengthen and improve privacy and security of individual users.

The study explores this angle, with a 'red team' approach, where a group of experts proactively identifies strengths and weaknesses in systems and organisations. We design AI/ML enabled methods to test and improve the resilience of IoT smart supply chains. We look at the challenges and potential for the use of privacy preserving AI/ML methods in regulatory red teams, such towards enabling data protection compliance. The paper builds upon the foundation of existing knowledge developed from three PETRAS projects (CRACS, 2018; IAM, 2018; P Radanliev, Nicolescu, De Roure, & Huth, 2019), but with a specific focus on Artificial Intelligence and Machine Learning (AI/ML) in IoT risk analytics. It benefits from the already established strong transformative and impactful research knowledge, but with a focus on the topic of securing the edge through AI/ML real time analytics. To avoid overlapping with earlier work, this article avoids many relevant areas that have been addressed in the working papers and project reports that can be found in pre-prints online (P. Radanliev, De Roure, Nicolescu, & Huth, 2019; P Radanliev, Roure, Nurse, & Nicolescu, 2019; Petar Radanliev, 2019a, 2019c, 2019b; Petar Radanliev, Charles De Roure, Nurse, Burnap, & Montalvo, 2019; Petar Radanliev et al., 2019, 2019, 2019, 2019; Petar Radanliev, De Roure, Nurse, Montalvo, & Burnap, 2019a, 2019b; Petar Radanliev, De Roure, Nurse, Montalvo, Burnap, et al., 2019; Petar Radanliev, De Roure, Nurse, Nicolescu, Huth, et al., 2019a, 2019c; Petar Radanliev, De Roure, Nurse, Burnap, Anthi, et al., 2019b). This working papers and project reports work enabled the cognition engine to be developed, tested and verified, though the active engagement with the user community and through responding to the new Internet of Things (IoT) risk and security developments as they emerged during the research. The novelty of this article is the relationship of this work to AI/ML and predictive analytics.

## 1.1 Motivation and methodology

A taxonomic approach is used for the evaluations of progress in the area of supply chain integration in the Industrial Internet of Things and the Industry 4.0, with a specific focus on the mitigation of cyber risks. An analytical framework is presented, based on a critical assessment with respect to issues related to new types of cyber risk and the integration of supply chains in new technologies. The approach is used to develop a transformational roadmap for the Industrial Internet of Things in Industry 4.0 supply chains of Small and Medium Enterprises (SMEs). The literature review includes 173 academic and industry papers and compares the academic literature with the established supply chain models. Taxonomic review is used to synthesise existing academic and practical research. Subsequently, case study research is applied to design a transformational roadmap. This is followed by the grounded theory methodology, to compound and generalise the findings into analytical framework. This results in a new analytical framework based, whereby articles are grouped followed by a series of case studies and vignettes and a grounded theory analysis.

The analytical framework drives the process of compounding knowledge from existing supply chain models and adapting the cumulative findings to the concept of supply chains in Industry 4.0. The findings from this study present a new approach for Small and Medium Sized companies to transform their operations in the Industrial Internet of Things and Industry 4.0. A supply chain is a system for moving products from supplier to customer and supply chain operational changes from digital technologies would specifically affect the small and medium sized companies (SMEs) because they lack the expertise, know-how, experiences and technological recourses of large enterprises (Petar Radanliev, 2014). A new approach for businesses and supply chain strategies is needed for the SME's to adapt to a changing environment. To build such approach, designing cases studies (Blatter & Haverland, 2012), with the ethnographic and discourse approaches to technology use and technology development is applied to the theory construction (David, 2005).

## 1.2 Our methodology

Methodologically, the article draws on a number of different sources and research methods, including a taxonomic review as a discourse of literature (Paltridge, 2017), case study research (Blatter & Haverland, 2012) including open and categorical coding, with discourse analysis and grounded theory. These methods are used in combination for conducting a systematic literature review. The data and the findings are synthesised using the grounded theory approach of categorising the emerging concepts (Glaser & Strauss, 1967). The case study research was performed on five I4.0 national initiatives and their technological trends in relation to IIoT product and services for a diverse set of industries. The diversity of the study participants represented in the sample population, is analysed with reference to the 'Industry Classification Benchmark' (FTSE Russell, 2018) to determine the industry representativeness in the selected I4.0 national initiatives and their technological trends.

To ensure validity of the conceptual system, the study applied qualitative research techniques (Easterby-Smith, Thorpe, & Lowe, 2002; Eriksson & Kovalainen, 2008; Gummesson, 2000), complimenting method for grounded theory (Charmaz, 2006), with open and categorical coding subsequently (Goulding, 2002). Discourse analysis is applied to evaluate and interpret the connotation behind the explicitly stated approaches (Eriksson & Kovalainen, 2008), along with tables of evidence (Eisenhardt, 1989) and conceptual maps (Miles, Huberman, & Saldaña, 1983).

## 1.3 Article roadmap

The sub-chapter 2.1 defines how SME's can integrate existing supply chain models; 2.2 defined the supply chain technical challenges from modern technological concepts; 2.3 defines how SME's can integrate cloud technologies into their supply chain management; 2.4 defines how SME's can integrate real-time IIoT technologies into their supply chain management; and 2.5 how SME can integrate cyber recovery planning into their supply chain management. Chapter 3 applies case study and grounded theory to categorise the I4.0 design principles. Chapter 4 presents the analytical framework and a transformational roadmap for integrating SMEs supply chains in the IIoT and I4.0.

## 2  Taxonomic review

The literature review covers a vast area of internet-of-things, cyber physical systems, industry 4.0, cyber security, and supply chain topics, e.g. digitisation, automation and autonomy. The literature review applies a taxonomic approach and follows the process of synthesising the most prominent categories, emerging from the reviewed literature. This follows the grounded theory approach of categorising emerging concepts (Glaser & Strauss, 1967). The emerging categories from the review are classified with open and categorical coding (Goulding, 2002) in the theory development chapter.

The taxonomic review of early supply chain models represents the foundation for our work on building the theoretical approach for integrating SME's in the Internet-of-Things and Industry 4.0. The focus of this review and the proposed approach is the Internet-of-Things approach within Supply Chain Management. Considering the vast literature on Supply Chain Management from decades of research, the review is focused on the key areas instead of covering too many topics. The review does not address the related areas of vertical and horizontal integration, smart supply chains, and supply chain visibility because that would represent too many topics and thereby lead to losing focus. Instead, the review applied presents an up-to-date literature review and categorises the best practices, design principles, common approaches, and standards affecting SME's supply chains in I4.0. This was considered as a relevant factor as many published models might rather apply to big corporations.

### 2.1  How to integrate existing supply chain models

Complexities remain in prioritising collective, as opposed to individual, performance improvement (Melnyk, Narasimhan, & DeCampos, 2014), and strategies commonly apply limited measurements (Van der Vaart & van Donk, 2008). Holistic design visualising how different types of integration creates different effects is proposed (Rosenzweig, Roth, & Dean, 2003). Thus, a hierarchical method can be applied for network design for deconstructing a complete supply chain that separates between the businesses and supply chain themes (Perez-Franco, 2016). This approach has never been applied for SME's designing I4.0 supply chains and its parameters will require altering to anticipate the similar and distinct features.

Following the taxonomic review method, the discourse of literature with open and categorical coding for discourse analysis and grounded theory, short summary of the areas is presented in the Table 1 outlining the design process on how SME's can integrate existing supply chain models. Along with the underlying factors driving the design (B. Lee et al., 2019b) in the digital age including aligning strategy with digital technology; implementations of Internet-enabled collaborative e-supply-chains; and integration of electronic supply chains. Table 1 details how to align and integrate existing supply chain models.

| How to integrate existing supply chain models | |
|---|---|
| Consensus on objectives | (Leng & Chen, 2012; Qu, Huang, Cung, & Mangione, 2010; Sakka, Millet, & Botta-Genoulaz, 2011) |
| Best level of integration | (Frohlich & Westbrook, 2001) |
| Organisational compatibility | (Mentzer et al., 2001) |
| Willingness to integrate operations | (Bryceson & Slaughter, 2010; Córdova, Durán, Sepúlveda, Fernández, & Rojas, 2012; Frohlich & Westbrook, 2001; Jayaram & Tan, 2010; Kaplan & Norton, 1996; Perez-Franco, 2016; Prajogo & Olhager, 2012; Sukati, Hamid, Baharun, & Yusoff, 2012) |
| Supply chain integration | (Al-Mudimigh, Zairi, & Ahmed, 2004; Frohlich & Westbrook, 2001; Manthou, Vlachopoulou, & Folinas, 2004; Vickery, Jayaram, Droge, & Calantone, 2003) |
| Individual integration obstacles | (Nikulin, Graziosi, Cascini, Araneda, & Minutolo, 2013) |
| Network design | (Dotoli, Fanti *, Meloni, & Zhou, 2005) |
| Supply chain hierarchical tree | (Qu et al., 2010) |
| Supply chain decomposition | (Schnetzler, Sennheiser, & Schönsleben, 2007) |
| Aligning strategy with digital technology | (W. Li, Liu, Belitski, Ghobadian, & O'Regan, 2016) |
| Internet-enabled collaborative e-supply-chains | (Pramatari, Evgeniou, & Doukidis, 2009) |
| Integration of electronic supply chains | (Yen, Farhoomand, & Ng, 2004) |

**Table 1: How to integrate existing supply chain models**

## 2.2 How to integrate modern technological concepts in supply chain management – technical challenges

The technical challenges for SME's integrating modern technological concepts, such as the I4.0 mostly evolve around the design challenges and the potential economic impact (loss) from cyber-attacks. But I4.0 also presents technical challenges in supply chains design and requires: software defined networks; software defined storage; protocols and enterprise grade cloud hosting; AI, machine learning, and data analytics; and mesh networks and peer-to-peer connectivity. The integration of such technologies in supply chains creates cyber security risk, for example from integrating less secured systems. Integrating the cyber element in manufacturing, also bring an inherent cyber risk. There are multiple attempts in literature where existing models are applied understand the economic impact of cyber risk. But there is no direct correlation between the higher cyber ranking and the industry application of digital infrastructure (Allen and Hamilton, 2014), thus challenges could be more related to performance metrics for security operations (Agyepong, Cherdantseva, Reinecke, & Burnap, 2019).

Building upon the taxonomic review method, the discourse of literature with open and categorical coding for discourse analysis and grounded theory, short summary is presented in the Table 2 outlining the technical challenges in the process of how to integrate modern technological concepts in supply chain management.

| How to integrate modern technological concepts in supply chain management – technical challenges | |
|---|---|
| Intelligent manufacturing equipment | (J. Lee, Bagheri, & Kao, 2015; Leitão, Colombo, & Karnouskos, 2016; Marwedel & Engel, 2016; Posada et al., 2015; Shafiq et al., 2015) |
| Machines capable of interacting with the physical world | (Brettel et al., 2016; Carruthers, 2016; Leonard, 2008; Lewis & Brigder, 2004; Marwedel & Engel, 2016; Rutter, 2015; L. Wang, 2013) |
| Software defined networks | (Kirkpatrick, 2013) |
| Software defined storage | (Ouyang et al., 2014) |
| Protocols and enterprise grade cloud hosting | (Carruthers, 2016) |
| AI, machine learning, and data analytics | (Kambatla, Kollias, Kumar, & Grama, 2014; Pan et al., 2015; Shafiq et al., 2015; Wan, Chen, Xia, Di, & Zhou, 2013) |
| Mesh networks and peer-to-peer connectivity | (Wark et al., 2007) |

| Understand the economic impact of cyber risk | (R. Anderson & Moore, 2006; Gordon & Loeb, 2002; Koch & Rodosek, 2016; Rodewald & Gus, 2005; Roumani, Fung, Rai, & Xie, 2016; Ruan, 2017; World Economic Forum, 2015) |
|---|---|
| Understanding the shared risk | (Koch & Rodosek, 2016; Rajkumar, Lee, Sha, & Stankovic, 2010; Ruan, 2017; Zhu, Rieger, & Basar, 2011) |
| Cyber risk estimated loss range | (Biener, Eling, & Wirfs, 2014; DiMase, Collier, Heffner, & Linkov, 2015; Koch & Rodosek, 2016; Ruan, 2017; Shackelford, 2016) |

**Table 2: How to integrate modern technological concepts in supply chain management – technical challenges**

### 2.3 How to integrate cloud technologies in supply chain management

To reduce costs and cyber risk, cloud technologies could enable value creation and value capture, through machine decision making (D. De Roure, Page, Radanliev, & Van Kleek, 2019). This would create service oriented planning (Akinrolabu, Nurse, Martin, & New, 2019). The social machines (D. De Roure et al., 2019) should be seen as the connection between physical and human networks (Shadbolt, O'Hara, De Roure, & Hall, 2019), operating as systems of systems (Boyes, Hallaq, Cunningham, & Watson, 2018), representing mechanisms for real-time feedback (David De Roure, Hooper, Page, Tarte, & Willcox, 2015) from users and markets (Marwedel & Engel, 2016).

Building upon the taxonomic review and the analytical framework based on taxonomic format, the Table 3 outlines a short summary of the design process for integrating cloud technologies into supply chain management.

| How to integrate cloud technologies in supply chain management | |
|---|---|
| Integrate cloud technologies | (Akinrolabu et al., 2019; Giordano, Spezzano, & Vinci, 2016; Ribeiro, Barata, & Ferreira, 2010; Shafiq et al., 2015; Thramboulidis, 2015; Wahlster et al., 2013) |
| Internet-based system and service platforms | (Dillon, Zhuge, Wu, Singh, & Chang, 2011; La & Kim, 2010; Wahlster et al., 2013; Wan, Cai, & Zhou, 2015; Weyer, Schmitt, Ohmer, & Gorecky, 2015) |
| IIoT processes and services | (Hussain, 2017; Stock & Seliger, 2016) |
| Industrial value chain | (Brettel et al., 2016; Hermann, Pentek, & Otto, 2016; S. Wang, Wan, Li, & Zhang, 2016) |
| Value creation and value capture | (Müller et al., 2018) |
| Machine decision making | (Evans & Annunziata, 2012; L. Wang, 2013) |
| Service oriented architecture | (La & Kim, 2010; L. Wang, Törngren, & Onori, 2015; Weyer et al., 2015) |
| Cloud distributed manufacturing planning | (Faller & Feldmüller, 2015; Posada et al., 2015) |
| Compiling of data, processes, devices and systems | (D. De Roure et al., 2019; Evans & Annunziata, 2012; Shafiq et al., 2015) |
| Model-driven manufacturing systems | (Jensen, Chang, & Lee, 2011; Shi, Wan, Yan, & Suo, 2011; L. Wang, Wang, Gao, & Váncza, 2014) |
| Model-based development platforms | (Ringert et al., 2015; Stojmenovic, 2014) |
| Social manufacturing | (Bauer et al., 2015; J. Lee, Kao, & Yang, 2014; Shadbolt et al., 2019; Wahlster et al., 2013; Wan et al., 2015) |
| Mechanisms for real-time distribution | (David De Roure et al., 2015; Kang, Kapitanova, & Son, 2012; Shi et al., 2011; Tan et al., 2008) |

**Table 3: How to integrate cloud technologies in supply chain management**

## 2.4 How to integrate modern technological concepts into supply chain management - real-time IIoT technologies

Digital supply chains should counteract components modified to enable a disruption. This could be supported by standardisation of design (J. Nurse, Creese, & De Roure, 2017) but risk assessing is still a key problem (Petar Radanliev et al., 2020). The reason for this is that digital cyber supply chain networks need to be: secure, vigilant, resilient and fully integrated (Craggs & Rashid, 2017) and encompass the security and privacy (Anthonysamy, Rashid, & Chitchyan, 2017).

The taxonomic review and the analytical framework in Table 4 outlines a short summary of the design process on how to integrate real-time IIoT technologies in supply chain management.

| **How to integrate real-time IIoT technologies in supply chain management** | |
|---|---|
| Real-time feedback from users and markets | (Hermann et al., 2016) |
| Information security for data in transit | (DiMase et al., 2015; Longstaff & Haimes, 2002; Toro, Barandiaran, & Posada, 2015) |
| Access control | (DiMase et al., 2015; Evans & Annunziata, 2012; Rajkumar et al., 2010) |
| Life cycle process | (Benveniste, 2010; Benveniste, Bouillard, & Caspi, 2010; Sokolov & Ivanov, 2015) |
| Counteract components | (DiMase et al., 2015; Evans & Annunziata, 2012) |
| Standardisation of design and process | (Ruan, 2017; Sangiovanni-Vincentelli, Damm, & Passerone, 2012) |
| Secure, vigilant, resilient and fully integrated | (Giordano et al., 2016) |
| Electronic and physical security of real-time data | (Almeida, Santos, & Oliveira, 2016; Niggemann et al., 2015) |

**Table 4: How to integrate real-time IIoT technologies in supply chain management**

## 2.5 How to integrate cyber recovery planning into supply chain management

The I4.0 brings inherent cyber risks and digital supply chains require cyber recovery plans supported with machine learning, enabling machines to perform autonomous decisions (Tanczer, Steenmans, Elsden, Blackstock, & Carr, 2018) and a design support system (B. Lee, Cooper, Hands, & Coulton, 2019a). To improve the response and recovery planning, digital supply chains need to be supported by feedback and control mechanisms, supervisory control of actions (Safa, Maple, Watson, & Von Solms, 2018). Most of these recommendations also apply to large enterprises. However, large enterprises have the recourses to control the entire supply chain, while SME's frequently have to integrate their supply chain operations (Petar Radanliev, 2015a, 2016). Integrating multiple SME's in the supply chain requires higher visibility and coordination between participants (Petar Radanliev, 2015b, 2015c).

Finally, the taxonomic review of literature and the analytical framework in Table 5 outlines a short summary of the design process on how SME's can integrate cyber recovery planning into their supply chain management.

| How to integrate cyber recovery planning in supply chain management | |
|---|---|
| Autonomous cognitive decisions | (Maple, Bradbury, Le, & Ghirardello, 2019; Niggemann et al., 2015; Pan et al., 2015; Wan et al., 2013) |

| Self-aware machines | (Weyer et al., 2015) |
|---|---|
| Self-optimising production systems | (Brettel et al., 2016; Posada et al., 2015; Shafiq et al., 2015; Wan et al., 2015) |
| Software assurance and application security | (Hussain, 2017; J. Lee et al., 2014; Niggemann et al., 2015) |
| Structured communications | (Almeida et al., 2016) |
| Cloud computing techniques | (Petrolo, Loscri, & Mitton, 2016) |
| Feedback and control mechanisms | (Niggemann et al., 2015) |
| Dynamics anti-malicious and anti-tamper control | (Benveniste, 2010; Sokolov & Ivanov, 2015) |

**Table 5: How to integrate cyber recovery planning in supply chain management**

## 2.6 The key gaps in the literature emerging from the taxonomic review of literature and the analytical framework

This review of technological trends on supply chain adoption confirms that SME's would benefit from a standardisation references for managing I4.0 complexities and IIoT resources efficiently. The key gaps in the literature which confirm that SMEs would benefit from standardisation reference are:

- Existing I4.0 architectures, lack clarification on designing individual components of I4.0 supply chains.
- The SME's need to integrate cloud technologies in their supply chains.
- The SME's digital supply chains need to encompass the security and privacy, along with electronic and physical security of real-time data.
- In the I4.0 supply chains, machines should connect and exchange information through cyber network and be capable of autonomous cognitive decisions.
- The SMEs need security measures to protect themselves from a range of attacks in their supply chains, while cyber attackers only need to identify the weakest links.
- The weakness of existing cyber risk impact assessment models is that the economic impact is calculated on organisations stand-alone risk, ignoring the impacts of sharing supply chain infrastructure.

The literature reviewed lacks clarification on the required design principles to address these gaps in individual levels of the I4.0 supply chains. Without such clarification, it is challenging to build a standardisation reference. In addition, supply chains design is still dominated by separation between established supply chain models, and the evolution of the IIoT. This separation is likely caused by the development of many established businesses and supply chain models before the rapid emergence of the IIoT.

## 3    Case study of five leading I4.0 technological trends

The gaps and key factors in current technological trends for I4.0 supply chain design integrating IIoT principles were derived from the taxonomic review. These are analysed through a case study of I4.0 frameworks in the current chapter. The case study specifically addresses the SME's needs for I4.0 know-how and develops a transformational roadmap of tasks and activities to reach a specific target state for the SME's supply chains. We have chosen to use a case study research-based methodology because it is recommended in recent literature for addressing the gaps in knowledge and for advancing the methodological rigour; this is done specifically by studying platforms on different architectural levels and in different industry settings (de Reuver, Sørensen, & Basole, 2017).

The case study design compares individual problems derived from the literature with the technological trends in industry today. Comparative analysis is applied which involves the five leading I4.0 initiatives and technological trends. The comparative analysis is building upon previous work on a comparison of 'Made in China 2025' and 'Industry 4.0' (L. Li, 2017), with an extended list of I4.0 initiatives. The justification for selecting the specific I4.0 initiatives was their richness in detail and explicitly stated strategies. The case study research initially reviewed 15 initiatives, worth mentioning, some countries have multiple I4.0 initiatives (e.g. USA, UK, Japan). But not all initiatives are discussed in great detail, as they lacked explicit details on I4.0 supply chains. The initial list of 15 initiatives reviewed are included in Table 6.

| I4.0 frameworks |
|---|
| **Germany** - Industrie 4.0 (GTAI, 2014). |
| **USA** - (1) Industrial Internet Consortium (IIC, 2017); (2) Advanced Manufacturing Partnership (AMP, 2013). |
| **UK** – (1) Catapults (John, 2017); (2) UK Digital Strategy (DCMS, 2017); (2) Made Smarter review 2017 (Siemens, 2017). |
| **Japan** - (1) Industrial Value Chain Initiative (IVI, 2017); (2) New Robot Strategy (NRS) (METI, 2015) and RRI (METIJ, 2015). |
| **France** - New France Industrial (NFI) – also known as: la Nouvelle France Industrielle or Industry of the Future (NIF, 2016) |
| **Nederland** - Smart Industry; or Factories of the Future 4.0 (Bouws et al., 2015). |
| **Belgium** - Made Different (Sirris and Agoria, 2017). |
| **Spain** - Industrie Conectada 4.0 (MEICA, 2015). |
| **Italy** - Fabbrica Intelligente (MIUR, 2014). |
| **China** - Made in China 2025 (SCPRC, 2017). |
| **G20** - New Industrial Revolution (NIR) (G20, 2016). |
| **Russia** - National Technology Initiative (NTI) (ASI, 2016). |

**Table 6: I4.0 frameworks reviewed**

The initiatives and their technological trends reviewed, embed the I4.0 and present a quick overview of the current state of the I4.0 supply chain adoption. The case study starts with the Industrial Internet Consortium (IIC, 2016), as the leading and most recent initiative, and follows with a case study of additional four I4.0 world leading initiatives.

The Industrial Internet Consortium (IIC) (IIC, 2016, 2017) promotes a fully connected and automated production line that brings the customer into the production process as a decision-maker. IIC supports highly automated (rules engines, protective overrides) and human operated (visualisation, intervention controls) usage environments.

The UK I4.0 report (DCMS, 2017) focuses extensively on the cloud integration in I4.0. While some initiatives are supported with direct examples of how the strategy can be executed (e.g. cloud data centres: Amazon, IBM, and Microsoft; or the cloud skills initiative to train public service in digital skills for development of cloud technology skills), other initiatives are not well defined. For example, the cloud-based software initiative states continuation towards common technology and lack a concrete action. This could in some instances be beneficial, as loosely defined standards provide flexibility in evolving as requirements change. Nevertheless, a concrete area of focus is required for the integration of SME's supply chains in I4.0. Another review report from the UK (Siemens, 2017) is focussed on industry rather than commerce. The report estimates a £185 billion value in the next ten years from four sectors construction, food and drink, pharmaceutical and aerospace sectors. The review makes three main recommendations for I4.0: adoption, innovation and leadership. While the value of this review cannot be denied, the claim of focus on industry can rather be described as the areas where government funding can help the industry. By reviewing the recommendations, it becomes clear that in each recommended area, public funding is required for achieving the goals. For example, the main areas (1) investing in a 'National Adoption Program'; (2) launching new innovation centres across the UK; (3) implementing large-scale digital transformational demonstrator programs and (4) pushing research and development in the identified areas; are all points that require public funding. Or the recommendation to up-skill a million industrial workers, again requires government funding. Even the seemingly leadership area of promoting the UK as a global pioneer in industrial digital technologies, which would fit in the government policy focus, is again confused with government subsidies as it calls for setting up a 'campaign', and setting up 'support implementation groups'. The objective of this article is to identify and categorise such policies and to present as industry led (and market focused) and not government led options for the UK and any other government that is aiming on developing their digital economies.

The most peculiar report is the Industrial Value Chain Initiative (IVI) (IVI, 2017). This I4.0 initiative, does not report any plans for real-time embedded systems or recovery plans. It is difficult to accept that Japan would miss out on these crucial principles for integrating IIoT in I4.0. It seems more likely that this initiative does not state such principles clearly in their reports. Nevertheless, a detailed review of all reports on the IVI (IVI, 2017) failed to identify any mentioning of real-time CPS or recovery plans.

The German initiative, Industrie 4.0 (GTAI, 2014; Industrie 4.0, 2017; Wahlster et al., 2013), covers the CPS and IIoT principles for cognitive evolution in I4.0. The German I.40 initiative promotes cloud computing integration with the Internet of Services, and proposes cloud-based security networks. The initiative states that automated real-time production is pioneered in Germany, but it does not specify with specific examples how real-time can be integrated in I4.0 and cognition is only mentioned, but not applied. The main criticism for Industrie 4.0 is that it does not state recovery plans.

In the case study, despite the lack of detail in the required categories, we include the Russian National Technology Initiative (NIT) (ASI, 2016) because of its significates in futuristic projections for I4.0 adoption. NIT represents more of a long-term forecasting for I4.0. The focus is on market network creations, and contributes with new insights to I4.0 by arguing that market creation for new technologies is the key to the future businesses and supply chain integration in I4.0. Similar argument that value capture processes should be focused on the ecosystem, is also present in literature (Metallo, Agrifoglio, Schiavone, & Mueller, 2018). But the forecasting does not address the issues of real-time cloud networks, and critically, does not provide recovery planning.

### 3.1   Categorising the I4.0 design principles emerging from the case study

These initiatives and their technological trends are applicable to SME's and to large enterprises. To identify the most prominent categories that apply to SME's supply chains, the comparative analysis applied the grounded theory approach to study and analyse the emerging trends and to organise into related categories and sub-categories. Through comparative analysis, a number of shortcomings in individual initiatives are identified. These shortcomings are addressed with the grounded theory design process of sub-categorising to the complimenting categories from the emerging I4.0 principles from the pre-selected 5 technological trends. More complicated problems emerge when the comparative analysis in Table 7 identifies that some of the national strategies propose very different approaches. The comparative analysis in Table 7 also identifies a number of gaps in national initiatives. By gaps, we refer to topics or a technological trends not incorporated in the associated national initiative.

| Supply chain integration - I4.0 | | | | |
|---|---|---|---|---|
| | Supply chain design - I4.0 | | | |
| I4.0 technological trends | Cloud integration of CPS and IIoT in I4.0 | Real-time CPS and IIoT in I4.0 | Autonomous cognitive decisions for CPS and IIoT in I4.0 | Recovery plans for CPS and IIoT in I4.0 |
| | | | | |
| IIC, 2016 | 1. Cloud-computing platforms. | 1. Adapt businesses and operational models in real time;<br>2. Customized product offers and marketing in real time. | 1. Fully connected and automated production line;<br>2. Support highly automated and human operated environments. | Gap - disaster recovery mentioned, but not incorporated. |
| DCMS, 2016 | 1. Cloud technology skills;<br>2. Cloud computing technologies;<br>3. Cloud data centres;<br>4. Cloud-based software;<br>5. Cloud-based computing;<br>6. Cloud guidance. | 1. Digital real-time and interoperable records;<br>2. Platform for real-time information. | 1. UK Robotics and Autonomous Systems;<br>2. Support for robotics and artificial intelligence;<br>3. Encourage automation of industrial processes;<br>4. Active Cyber Defence. | Gap |
| IVI, 2017 | 1. Cloud enabled monitoring;<br>2. Integration framework in cloud computing. | Gap | 1. Factory Automation Suppliers and IT vendors;<br>2. Utilisation of Robot Program Assets by CPS.s | Gap |
| Industrie 4.0 | 1. CPS automated systems;<br>2. Automated conservation of resources. | 1. Cloud computing;<br>2. Cloud-based security networks. | 1. Automated production;<br>2. Automated conservation of recourses. | Gap |
| NTI, 2015 | Gap | Gap | 1. Artificial intelligence and control systems | Gap |

**Table 7: Design principles emerging from the case study**

To resolve these gaps, individual areas are used as reference categories for building the analytical framework (which is presented later in Figure 3) that relates various areas and eliminates conflicts in different and sometimes contrasting I4.0 approaches. Following the grounded theory approach (Glaser & Strauss, 1967), the main categories of each individual initiative are separated into subcategories in Table 2 according to the gaps in their design principles.

## 4 Analytical framework and a transformational roadmap

The analytical framework development builds upon the taxonomic review of literature and starts with organising the most prominent categories of emerging approaches in literature. This process of organising concepts into categories, follows the grounded theory approach (Glaser & Strauss, 1967) and the open and categorical coding approach (Goulding, 2002). Discourse analysis is applied to evaluate and interpret the meanings behind the categories (Eriksson & Kovalainen, 2008), supported with tables of evidence (Eisenhardt, 1989) and conceptual diagrams (Miles et al., 1983) to present graphical analysis. The methodological approach is described in more details in Chapter 3 and in this chapter is focused on enabling SME's practitioners to identify the value of the proposed theoretical concept. The process of interpreting the connotation behind the categories, the tables of evidence and the conceptual diagrams are aimed specifically to present methodological approach with graphical analysis for SME's practitioners, as they normally need rather hands-on recommendations.

### 4.1 Pursuit of theoretical validity through case study research

In pursuit of theoretical validity, the methodological approach with graphical analysis was presented on the case study group discussions with experts from industry. The case study design primarily contributed to the process of identifying a hierarchical organisation of the methodological approach.

The graphical analysis was used as a tool during the group discussions to verify the themes, categories and subcategories and their hierarchical relationships. The group discussions included two different centres from Fujitsu: Artificial Intelligence and Coelition; and four different Cisco research centres: First Centre: Security and Trust Organisation, Second Centre: Advanced Services, Third Centre: Security Business Group, Fourth Centre: Cisco Research Centre. For the group discussion, the study recruited 20 experts and distinguished engineers. This approach to pursuing validity follows recommendation from existing literature on this topics (Axon, Alahmadi, Nurse, Goldsmith, & Creese, 2018; Eggenschwiler, Agrafiotis, & Nurse, 2016; Müller et al., 2018). The methodological approach advances conceptual clarity and provides clear definitions that specify the unit of analysis for digital platforms. These are identified as recommended areas for further research in recent literature (de Reuver et al., 2017).

## 4.2 Design principles for I4.0 supply chains

We place an emphasis on a cognitive I4.0 analytical framework. A cognitive I4.0 framework refers to the trend of automation, introduced by computing devices that are reasoning and making supply chain decisions for humans. The emerging applications and technologies are presented in the form of a grouping diagram (Figure 1) to visualise the required concepts for the integration of SME's supply chains in I4.0.

The grouping of concepts starts with the most prominent categories emerging from the taxonomy of literature: (1) self-maintaining machine connection for acquiring data and selecting sensors; (2) self-awareness algorithms for conversion of data into information (Ghirardello, Maple, Ng, & Kearney, 2018); (3) connecting machines to create self-comparing cyber network that can predict future machine behaviour (E. Anthi, Williams, & Burnap, 2018); (4) generates cognitive knowledge of the system to self-predict and self-optimise, before transferring knowledge to the user (Madaan et al., 2018); (5) configuration feedback and supervisory control from cyber space to physical space, allowing machines to self-configure, self-organise and be self-adaptive (J. Lee et al., 2015).

Following the methodology for reliable representation of the data collected, open coding is applied (Goulding, 2002) to the emerging categories for recovery planning in Figure 1. The conceptual diagram in Figure 1 present graphical analysis of the emerging design principles for cognition in IIoT digital supply chains. The emerging design principles in the conceptual diagram, also address the recommended gaps in recent literature on advancing methodological rigour by employing design research and visualisation techniques (de Reuver et al., 2017), such as the graphical analysis in the figure. The elements in the diagram emerge from the reviewed I4.0 technological trends, national initiatives and frameworks reviewed (Table 1) and the links between the elements emerge from the design principles identified in the case study (Table 2) for SME's supply chains in I4.0.

The findings in Figure 1 present the first stage of designing a dynamic and self-adapting system supported with artificial intelligence and Machine Learning (AI/ML) and real-time intelligence for predictive cyber risk analytics (PETRAS, 2020).

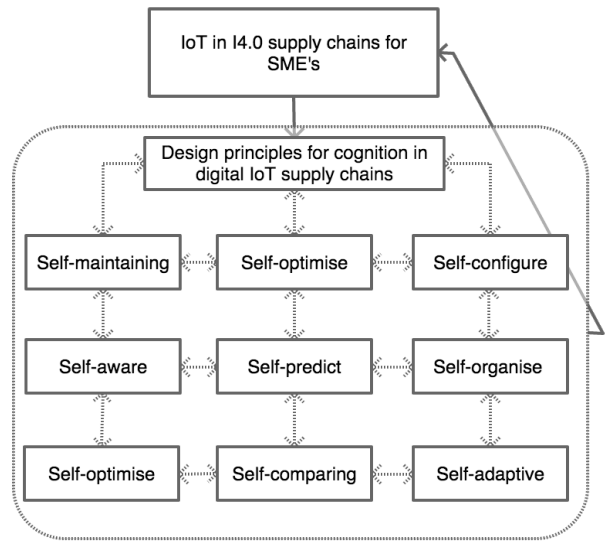

**Figure 1: Iterative learning and improvement in design principles – synthesised from the taxonomic review**

The described principles represent the beginning of a cognitive architecture for I4.0 supply chains. Such cognitive architecture allows for learning algorithms and technologies to be changed quickly and re-used on different platforms (Brettel et al., 2016; Niggemann et al., 2015), for creating multi-vendor production systems (Weyer et al., 2015) which is necessary for the I4.0 supply chains. A cognitive production systems would provide real-time synchronised coexistence of the virtual and physical dimensions (Shafiq et al., 2015).

The emergence of cognition, confirms that I4.0 supply chain design requires multi-discipline testing and verification (Balaji, Al Faruque, Dutt, Gupta, & Agarwal, 2015), including understanding of system sociology (Dombrowski & Wagner, 2014), because it operates in a similar method with social networks (Bauer et al., 2015; Wan et al., 2015). In the I4.0 supply chains, machines should connect and exchange information through networks (Toro et al., 2015) providing optimised production and inventory management (J. Lee et al., 2015; Wan et al., 2015; Weyer et al., 2015), and CPS lean production (Kolberg & Zühlke, 2015).

### 4.3 Cognitive architecture principles for recovery planning in I4.0 supply chains

I4.0 is expected to evolve from the traditional supply chain network into digital supply chain networks (Taylor, P., Allpress, S., Carr, M., Lupu, E., Norton, J., Smith et al., 2018). For digital supply chains to be considered secure and to ensure digital recovery planning is adequate, the supply chains need to be self-aware (P Radanliev et al., 2019), because a single failure could trigger a complex cascading effect, creating wide-spread failure (Breza, Tomic, & McCann, 2018).

Adding to this, distributed energy resource technologies such as wind power, create additional stress and vulnerabilities (Ahmed, Kim, & Kim, 2013; Marwedel & Engel, 2016). To ensure supply chains to be considered secure and to ensure digital recovery planning is adequate, advanced power electronics and energy storage are required for coordination and interactions (Leitão et al., 2016; Marwedel & Engel, 2016; Rajkumar et al., 2010), as well as physical critical infrastructure with preventive and self-correcting maintenance (Brettel et al., 2016; Leitão et al., 2016; Zhu et al., 2011).

Following the methodology for recognising the profounder concepts in the data (Goulding, 2002), categorical coding is applied as a complimenting method for grounded theory (Charmaz, 2006) to compare the emerging categories for recovery planning with the categories in the taxonomic review. In this process, discourse analysis is applied to interpret the data (Eriksson & Kovalainen, 2008), behind the explicitly stated categories in the taxonomic review, resulting in explicitly stated categories for recovery planning in Figure 2. The links between the elements in Figure 2 emerge from applying the grounded theory approach to relate the findings from the literature with the reviewed I4.0 technological trends, national initiatives and frameworks reviewed (Table 1) and the links between the elements as confirmed in the design principles (Table 7) and presented in Figure 1.

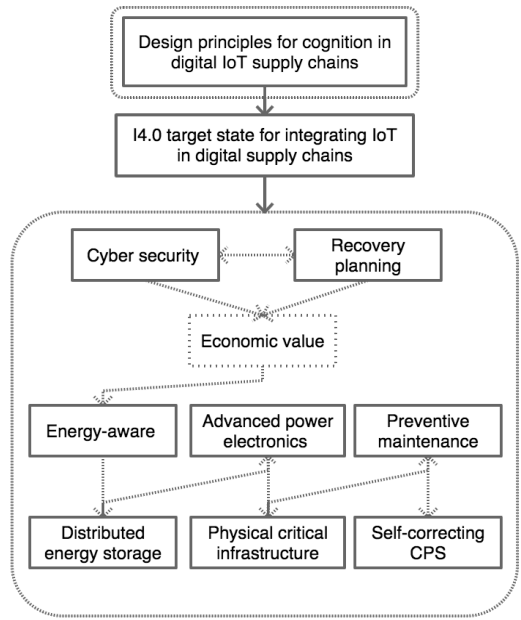

**Figure 2: I4.0 target state for integrating IIoT in digital supply chains**

The conceptual diagram in Figure 2 provides SME's with a bird's eye view of an I4.0 target state for integrating IIoT in SME's digital supply chains. The target state diagram advances an earlier approaches (Shaw, Snowdon, Holland, Kawalek, & Warboys, 2004) and presents the smart capability functions at a strategic, business process and technical level. This presents the second stage of designing a dynamic and self-adapting system supported with artificial intelligence and Machine Learning (AI/ML) and real-time intelligence for predictive cyber risk analytics (PETRAS, 2020). This will enhance capacities and assist in the creation of a comprehensive and systematic understanding of the opportunities and threats that arise when edge computing nodes are deployed, and when AI/ML technologies are migrated to the periphery of the internet and into local IoT networks.

### 4.4 Challenges for IIoT integration in Industry 4.0 supply chains

Apart from recovery planning, other challenges found in literature for SME's integration in Industry 4.0 supply chains are:

a) robustness, safety, and security (Akinrolabu et al., 2019; I. Brass, Tanczer, Carr, Elsden, & Blackstock, 2018; Irina Brass, Pothong, Tanczer, & Carr, 2019; Hahn, Ashok, Sridhar, & Govindarasu, 2013; Nicolescu, Huth, Radanliev, & De Roure, 2018a; Zhu et al., 2011);

b) control and hybrid systems (Agyepong et al., 2019; Leitão et al., 2016; J. R. Nurse, Radanliev, Creese, & De Roure, 2018; Shi et al., 2011);

c) computational abstractions (Ani, Watson, Nurse, Cook, & Maple, 2019; Madakam, Ramaswamy, & Tripathi, 2015; Petar Radanliev, De Roure, Nicolescu, et al., 2018; Rajkumar et al., 2010; Wahlster et al., 2013);

d) real-time embedded systems abstractions (Ghirardello et al., 2018; Kang et al., 2012; Leitão et al., 2016; Marwedel & Engel, 2016; PETRAS, 2020; Shi et al., 2011; Tan et al., 2008);

e) model-based development (Bhave, Krogh, Garlan, & Schmerl, 2011; Jensen et al., 2011; Rajkumar et al., 2010; Shi et al., 2011; Taylor, P., Allpress, S., Carr, M., Lupu, E., Norton, J., Smith et al., 2018; Wahlster et al., 2013); and

f) education and training (Faller & Feldmüller, 2015; Nicolescu, Huth, Radanliev, & De Roure, 2018b; Petar Radanliev et al., 2020; Rajkumar et al., 2010; Wahlster et al., 2013).

These challenges present the difficulties SME's face. SME's need protection across a range of new technologies, while attackers only need to identify the weak links (Eirini Anthi, Williams, Slowinska, Theodorakopoulos, & Burnap, 2019; Van Kleek et al., 2018). This reemphasises the need for recovery plans, which is not explicitly covered in the I4.0 initiates from the case study.

## 4.5 Future technologies for SME's integration in Industry 4.0 supply chains

Finally, the SME's need to plan for the adoption of future technologies, to reduce cost and ensure compliance with technological updates in their supply chain. Future technologies include the deployment of self-sustaining networked sensors (Rajkumar et al., 2010) and Cloud centric supply chains (Gubbi, Buyya, Marusic, & Palaniswami, 2013), symbiotic with the physical environment (Pan et al., 2015), creating eco-industrial by-product synergies (Pan et al., 2015; Stock & Seliger, 2016). Such supply chains would be supported with self-adapting distributed integrated-decentralised architecture (Stojmenovic, 2014; Wan et al., 2015), enabling applications to self-adjust and self-optimise own performance (Brettel et al., 2016; Shafiq et al., 2015). Where individual contract-based design is applied before platform-based design (Sangiovanni-Vincentelli et al., 2012), enabling multiple models of computation to act as a single system (Benveniste et al., 2010; Bhave et al., 2011).

### 4.6 Transformational roadmap for SME's supply chain design in I4.0

Here, we propose a transformational roadmap (Figure3), where individual concepts describe larger blocks of the I4.0 supply chains. The design initiates with applying the categories and sub-categories from the taxonomy and the emerging standards from the case study that are affecting SMEs supply chains in the I4.0 (Table 2). Then applying the grounded theory approach and following the recommendations from the literature reviewed, to relate the most prominent categories and its related subcategories into conceptual diagrams. This design processes integrates the categories and captures the best practices in industry. This methodological design process follows recommendations from literature (Strader, Lin, & Shaw, 1999), and shows how individual components can be integrated into an information infrastructure, with the technologies that can fit within the proposed transformational roadmap.

The synthesised categories and sub-categories in the transformational roadmap are related to the gaps from the taxonomic review. For instance, the categories emerging from the taxonomic review, and compounded to address the identified gap, before being hierarchically structured and organised in a step by step method. The transformational roadmap embodies a process of supply chain design decomposition, starting with a bird eye view of the synthesised models on businesses and supply chain design. Followed by the synthesised knowledge from the taxonomic review and the case study, embodied to SMEs supply chains in the I4.0. The transformational roadmap design in Figure 3 embodies a process of ideas and concepts conceived as an interrelated, interworking set of objectives and applies directive, conventional and summative analysis to relate the recovery planning with the design categories. The transformational roadmap design integrates the findings from literature review on recovery planning, with the findings from the case study and relates recovery planning with principles represented in the categories for SME's supply chain networks in I4.0.

The principles for SME's supply chain networks in I4.0 supply chains are related to the findings and the gaps identified in the taxonomic review of the earlier supply chain integration models before I4.0. The findings are specifically related to advancing and generalising the previous case specific work on the implementations of Internet-enabled collaborative e-supply-chain initiatives (Pramatari et al., 2009) and integrated electronic supply chains (Yen et al., 2004). Then the findings and the gaps identified in the case study of the I4.0 initiatives and their technological trends (e.g. that recovery plans are not explicitly provided in such initiatives) are addressed with specific action objectives from the taxonomic review.

The logic behind the steps in Figure 3 represents the current understanding of the academic and industry papers and publications reviewed in this article. The choice and sequence of steps is supported by the taxonomic review in chapter 3 and the analysis of the I4.0 technological trends, national initiatives and frameworks in chapter 4. The rationale as to why the particular steps and their proposed sequence are chosen derive from the design principles in Figure 1 and the target state in Figure 2. In addition, the transformational roadmap in Figure 3 encompasses material and understanding derived from review and analysis of 173 academic and industry papers, analysed with the grounded theory approach to ensure the work is repeatable and is verified with the rigour of a time tested and established method for conducting a systematic review of literature.

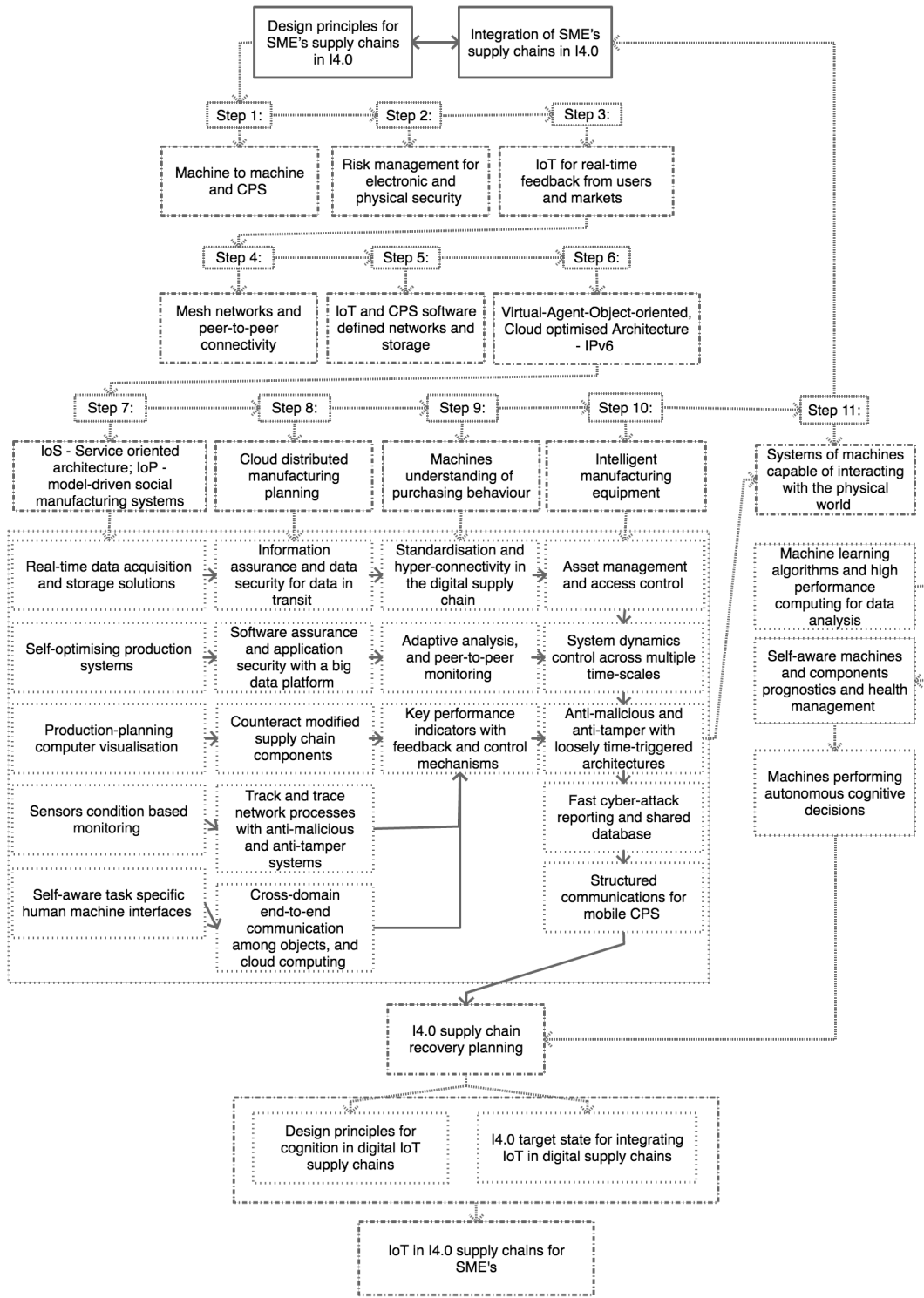

**Figure 3: Analytical framework based on taxonomic/cladistic format: transformational roadmap for supply chain integration in I4.0.**

The transformational roadmap in Figure 3 evaluates the relationship between the IIoT technological trends and derives with a process of digitalising SME's supply chain. The transformational roadmap recommends the development of cognitive supply chain principles that enable storing and sharing knowledge. This is of specific relevance to SME's because SMEs and large enterprises do not have the same recourse and using existing knowledge enhances the I4.0 adaptation process in SME's. Figure 3 presents the final stage of the conceptual designing a dynamic and self-adapting system supported with artificial intelligence and Machine Learning (AI/ML) and real-time intelligence for predictive cyber risk analytics (PETRAS, 2020). By integrating AI/ML in the risk analytics, we devise a new approach for cognitive data analytics, creating a stronger resilience of systems through cognition in their physical, digital and social dimensions. It is expected that Web Science will be increasingly more present in the physical world because of smart and connected devices (David De Roure et al., 2019). Our approach resolves around understanding how and when such connections causes compromises to happen, and to enable systems to adapt and continue to operate safely and securely when they have been compromised. Cognition through AI/ML is the key topic of this research and cognitive real time intelligence would enable systems to recover and become more robust.

The transformational roadmap structures the principles for recovery planning in SME's digital supply chains. The principles present the explicit relationships derived from the taxonomies and the case study. The explicit relationships between the principles for recovery planning in cognitive IIoT supply chain networks, enables the assessment of individual technical risk for a given vulnerability. Through a visualisation of the explicit relationships in digital SME's supply chains, the technical risk for a given vulnerability can be better assessed, e.g. by applying the Common Vulnerability Scoring System (CVSS) (CVSS, 2019).

The analytical framework also considers the issues with adoption, as it seems that in most of the reviewed literature everyone tries to create their own model. The taxonomic review and the case study identified the gaps in existing models, and the transformational roadmap made the solutions visible in an explicit format. The transformational roadmap in this paper, however, is dependent on given vulnerability being assessed by existing cyber risk assessment models (e.g. CVSS, 3.1) and analysed with existing cyber risk analysis models (e.g. FAIR-U tool). Hence, the analytical framework is promoting the development of a generally accepted cyber security framework; this is also called for in current research work (FAIR, 2020). The analytical framework represents a generic reusable approach, to be used by SME's for supply chain strategy development for I4.0 by supply chain stakeholders and practitioners.

The analytical framework in Figure 3 connects the supply chains and the impact of cyber risk to human-computer interactions in different supply chain management systems with artificial intelligence. This can provide supply chain predictive feedback sensors. These feedback sensors would represent dynamic real time data mechanisms that assist and enable better understanding of the problem - prior to cyber-attacks. The reliability of cyber risk impact assessments could increase significantly if decisionmakers have a dynamic and self-adopting AI enhanced feedback sensors to assess, predict, analyse and address the risks of cyber-attacks in the supply chain.

The analytical framework in Figure 3 firstly identifies and articulates some of the possible supply chain solutions for the role of machine learning (ML) in designing dynamic automated predictive feedback cognitive system, supported with real-time intelligence. Secondly, the analytical framework in Figure 3 identifies cyber risk analytic approaches with dynamic real-time and ML self-adapting enhanced technologies that enable predictive risk analytics.

In doing this work we are acutely aware that adding automation and further coupling to a distributed system also brings new opportunities for cascading effects and exposing new attack surfaces. These concerns are fundamental to the cognition engine design, especially in the areas with increased automation of processes which have classically required human interaction.

Furthermore, in terms of the (un)availability of data, lessons can be learned from previous research on data strategies (Petar Radanliev, De Roure, Nurse, Montalvo, & Burnap, 2019b). The volume of data generated creates diverse challenges for developing data strategies in a variety of verticals (ex. AI/ML, ethics, business requirements). Simultaneously, designing a supply chain cyber security architecture for complex coupled systems, while understanding the impact, demands data strategy optimisation and decision making on collecting and assessment of probabilistic data when edge computing nodes are deployed, presents a socio-technical research problem. The research is also strongly related to  personal perceptions of risk because of collecting probabilistic data at the edge interact with data regulations, standards and policies. These data perceptions, regulations and policies are strongly considered in our approach for integrating ML in supply chain cyber risk data analytics. A cybersecurity architecture for impact assessment with ML cyber risk analytics must meet public acceptability, security standards, and legal scrutiny. With consideration of the above, the research integrated areas such as impact, policy and governance recommendations, while continuously anticipating aspects of computer science to develop and design architectures for ML in supply chain cyber risk data analytics. The research contributes to knowledge by integrating supply chain management with ML and cyber risk analytics that have not been previously integrated in a research on securing supply chains, and thus promote the field of developing a dynamic and self-adopting methodology to assess, predict, analyse and address the risks of cyber-attacks in the supply chains.

## 4.7 Discussion and main findings

The study applies taxonomic review and case study research to derive with the design principles for a analytical framework with a transformational roadmap that enables the process of integrating SME's business and supply chains in the I4.0 network. The analytical framework captures the best practices in industry, and defines the differences and similarities between I4.0 technological trends. Major projects on I4.0 are reviewed to present the landscape for cutting edge developments in IIoT, offering us a comprehensive picture of the current state of supply chain adoption.

The analytical framework and the transformational roadmap do not address the aspect of people but instead the focus is on the process aspects of Industry 4.0. While the people aspects are important given the general shortage of individuals with appropriate digital skills, this problem has been addressed by some countries e.g. Australia with a points-based system for attracting people with appropriate digital skills. The process aspects were determined as more important because Industry 4.0 is going to require changes in business practices (and hence processes), and there are multiple approaches to structuring such processes as identified in the case study of I4.0 initiatives. Creating a unified approach to process, with a step-by-step transformational roadmap was missing in academic and industry literature. The design principles in Figure 1, the target state in Figure 2 and the transformational roadmap in Figure 3 derive from the analysis of the state-of-the-art literature and the leading I4.0 initiatives, presenting a unified approach to process development.

### 4.7.1 Main findings pertaining to the analytical framework

***Standardisation reference for I4.0 supply chains***

The I4.0 adoption pertains:

a) Standardisation reference architecture (Ahmed et al., 2013; Petar Radanliev et al., 2020; Stock & Seliger, 2016; Wahlster et al., 2013; Weyer et al., 2015).

b) Existing I4.0 architectures (Giordano et al., 2016; Hermann et al., 2016; J. Lee et al., 2015), lack clarification on designing individual components of I4.0 supply chains.

c) Despite the strong interest in literature and industry for designing I4.0 and cyber risk standardisation reference architectures, this is the first attempt to integrate various academic models with industry and government initiatives.

The design principles of the analytical framework demystify this, by comparing models from academic literature with major projects from industry/governments and clarify individual levels of I4.0 supply chain design.

***Cloud integration of CPS and IIoT of SME's in the I4.0 supply chains***

The SME's need to:

d) Integrate cloud technologies in their supply chains (Giordano et al., 2016; Ribeiro et al., 2010; Shafiq et al., 2015; Thramboulidis, 2015; Wahlster et al., 2013).

This study derives with the determining factors for an IIoT approach within Supply Chain Management in I4.0, with the focus on SME's cloud technologies. Some of the direct recommendations in the design principals include the deployment of self-sustaining networked sensors and Cloud centric supply chains, symbiotic with the physical environment.

***Real-time CPS and IIoT in I4.0***

The SME's digital supply chains need to:

e) Encompass the security and privacy (Anthonysamy et al., 2017), along with electronic and physical security of real-time data (Agyepong et al., 2019).

The findings from this study enable SMEs to integrate IIoT in their I4.0 businesses and supply chains with a step-by-step transformational roadmap. The transformational roadmap includes the design principles and outlines the process for integrating SME's with real-time enabled IIoT in the I4.0 supply chains.

*Autonomous cognitive decisions for CPS and IIoT in I4.0*

In the I4.0 supply chains, machines should:

f) Connect and exchange information through cyber network and be capable of autonomous cognitive decisions (Kolberg & Zühlke, 2015; J. Lee et al., 2015; Toro et al., 2015; Wan et al., 2015; Weyer et al., 2015).

Existing literature lacks clarification on how such automation can be designed in the context of I4.0 supply chains. The study derives with design principles for cognition in digital IIoT supply chains and an I4.0 target state for integrating IIoT in digital supply chains.

*Cyber risk concerns*

The SMEs need security measures to protect themselves from a range of attacks in their supply chains, while cyber attackers only need to identify the weakest links. Hence, the cyber risk creates a disadvantage for SMEs as they need to invest a great deal of resources into cyber protection and recovery planning. The transformational roadmap enables SME's to visualise and charts them on the path to beginning to address the cyber risk. While SMEs need to embrace the I4.0 in their supply chains, but SMEs also need to be aware of the inherent cyber risks. The taxonomic review and the case study in this study, emphasised the vast areas of cyber risk and brought the attention on cyber recovery.

*Cyber risk assessment problems*

The weakness of existing cyber risk impact assessment models is that the economic impact is calculated on organisations stand-alone risk, ignoring the impacts of sharing supply chain infrastructure (J. Nurse et al., 2017; Petar Radanliev, De Roure, Cannady, et al., 2018; Petar Radanliev, De Roure, Nurse, et al., 2018). In addition, there is an inconsistency in measuring supply chain cyber risks, which is caused by the lack of understanding of supply chain operations in I4.0. This study enables the process of visualising the shared risk in supply chains. The visualisation of such risks is vital for calculating and planning for the impact to the SMEs operating in the I4.0.

### *Recovery plans for CPS and IIoT in I4.0*

Clarity on disaster recovery plans is missing in all of the I4.0 technological trends analysed in the case study, with no explanation on details or on how recovery planning would be executed. This is of concern as in the literature the recovery planning is strongly emphasised. The analytical framework derives with direct recommendations that would improve the response and recovery planning in the SME's supply chains. Some of the recommendations include the need for feedback and control mechanisms, supervisory control of actions, and dynamics anti-malicious and anti-tamper control.

## 5     Conclusion

By integrating AI/ML in the risk analytics, in this article we devise a new approach for cognitive data analytics, creating a stronger resilience of systems in their physical, digital and social dimensions. Our approach resolves around understanding how and when compromises happen, to enable systems to adapt and continue to operate safely and securely when they have been compromised. Cognition through AI/ML is the key topic of this research and cognitive real time intelligence would enable systems to recover and become more robust.

This paper identifies a dynamic and self-adapting system supported with AI/ML and real-time intelligence for predictive cyber risk analytics. This will enhance national capacities and assist in the creation of a comprehensive and systematic understanding of the opportunities and threats that arise when edge computing nodes are deployed, and when AI/ML technologies are migrated to the periphery of the internet and into local IoT networks.

We used a series of new design principles to derive a transformational roadmap and a new analytical framework for the SME's supply chains integration in I4.0. Despite the strong interest in the value for SME's supply chain from IIoT and I4.0, this research represents the first attempt to synthesise and compare knowledge from literature with expert's opinions. This knowledge was applied to design a step by step approach for the SME's supply chains integration with IIoT technologies in the I4.0. In the design process, the SME's supply chain networks are related to the I4.0 initiatives and their technological trends.

The research discovered that successful adaptation of IIoT technologies, depends largely on the cyber recourses. This specifically concerns SME's as they do not have the same supply chain recourses as large enterprises. The new design enables SME's to visualise the required cyber resources and the integration process and the transformational roadmap the integration process of IIoT technologies consolidated in the cyber themes of the future makeup of supply chains. The analytical framework can also be applied to visualise and assess their exposure to cyber risk and to design cyber recovery. This visualisation also supports policy development by decomposing operational system with concrete and workable action plans, that would transition the economic and social systems towards new cyber capabilities.

At a higher analytical level, the article presents new design principles, a transformational roadmap and a new analytical framework, for small and medium enterprises to approach the new supply chains technological challenges in industry 4.0. The research's insights are based on a literature analysis, case study research and a grounded theory methodology. The validation of these research results was checked with experts from two corporations, Cisco Systems and Fujitsu. The case study is also informed by the sustained engagement of the UK EPSRC IIoT Research Hub PETRAS[i] with a broad set of user partners for a wide range of private sectors, government agencies, and charities at international scale.

**Declaration:**

**Availability of data and material:** Not applicable.

**Acknowledgements:** Eternal gratitude to the Fulbright Program.

**Competing interests' section:** The authors declare that they have no competing interests.

**Funding sources:** This work was funded by the UK EPSRC [with the PETRAS 2 projects: RETCON and CRatE, grant number: EP/S035362/1, EP/N023013/1, EP/N02334X/1] and by the Cisco Research Centre [grant number 1525381]. Working papers and project reports prepared for the PETRAS National Centre of Excellence and the Cisco Research Centre can be found in pre-prints online.

# 6 References

Agyepong, E., Cherdantseva, Y., Reinecke, P., & Burnap, P. (2019). Challenges and performance metrics for security operations center analysts: a systematic review. *Journal of Cyber Security Technology*, *4*(1), 1–28. https://doi.org/10.1080/23742917.2019.1698178

Ahmed, S. H., Kim, G., & Kim, D. (2013). Cyber Physical System: Architecture, applications and research challenges. *2013 IFIP Wireless Days (WD)*, 1–5. https://doi.org/10.1109/WD.2013.6686528

Akinrolabu, O., Nurse, J. R. C., Martin, A., & New, S. (2019, November 1). Cyber risk assessment in cloud provider environments: Current models and future needs. *Computers and Security*, Vol. 87, p. 101600. https://doi.org/10.1016/j.cose.2019.101600

Al-Mudimigh, A. S., Zairi, M., & Ahmed, A. M. M. (2004). Extending the concept of supply chain:: The effective management of value chains. *International Journal of Production Economics*, 87(3), 309–320.

Allen and Hamilton. (2014). *Cyber Power Index: Findings and Methodology*. Retrieved from https://www.sbs.ox.ac.uk/cybersecurity-capacity/system/files/EIU - Cyber Power Index Findings and Methodology.pdf

Almeida, L., Santos, F., & Oliveira, L. (2016). *Structuring Communications for Mobile Cyber-Physical Systems*. https://doi.org/10.1007/978-3-319-26869-9_3

AMP. (2013). Advanced Manufacturing Partnership. In *NIST Advanced Manufacturing Office*. Retrieved from https://www.nist.gov/amo/programs

Anderson, G. (2016). The Economic Impact of Technology Infrastructure for Smart Manufacturing. *NIST Economic Analysis Briefs*, 4. https://doi.org/10.6028/NIST.EAB.4

Anderson, R., & Moore, T. (2006). The Economics of Information Security. *Science AAAS*, 314(5799), 610–613. Retrieved from http://science.sciencemag.org/content/314/5799/610.full

Ani, U. D., Watson, J. D. M., Nurse, J. R. C., Cook, A., & Maple, C. (2019). A Review of Critical Infrastructure Protection Approaches: Improving Security through Responsiveness to the Dynamic Modelling Landscape. *PETRAS/IET Conference Living in the Internet of Things: Cybersecurity of the IoT - 2019*, 1–16. Retrieved from http://arxiv.org/abs/1904.01551

Anthi, E., Williams, L., & Burnap, P. (2018). Pulse: an adaptive intrusion detection for the internet of things. *Living in the Internet of Things: Cybersecurity of the IoT*, 35 (4 pp.). https://doi.org/10.1049/cp.2018.0035

Anthi, Eirini, Williams, L., Slowinska, M., Theodorakopoulos, G., & Burnap, P. (2019). A Supervised Intrusion Detection System for Smart Home IoT Devices. *IEEE Internet of Things Journal*, 6(5), 9042–9053. https://doi.org/10.1109/JIOT.2019.2926365

Anthonysamy, P., Rashid, A., & Chitchyan, R. (2017). Privacy Requirements: Present & Future. *2017 IEEE/ACM 39th International Conference on Software Engineering: Software Engineering in Society Track (ICSE-SEIS)*, 13–22. https://doi.org/10.1109/ICSE-SEIS.2017.3

Ashton, K. (2011). In the real world, things matter more than ideas. *RFID Journal*, 22(7). Retrieved from http://www.rfidjournal.com/articles/pdf?4986

ASI, A. for strategic initiatives. (2016). National Technology initiative, Agency for Strategic Initiatives. Retrieved May 10, 2017, from Government of Russia website: https://asi.ru/eng/nti/

Axon, L., Alahmadi, B., Nurse, J. R. C., Goldsmith, M., & Creese, S. (2018). Sonification in Security Operations Centres: What do Security Practitioners Think? *Proceedings of the Workshop on Usable Security (USEC) at the Network and Distributed System Security (NDSS) Symposium*, 1–12. Retrieved from https://www.cs.ox.ac.uk/files/9802/2018-USEC-NDSS-aangc-preprint.pdf

Balaji, B., Al Faruque, M. A., Dutt, N., Gupta, R., & Agarwal, Y. (2015). Models, abstractions, and architectures. *Proceedings of the 52nd Annual Design Automation Conference on - DAC '15*, 1–6. https://doi.org/10.1145/2744769.2747936

Bauer, W., Hämmerle, M., Schlund, S., & Vocke, C. (2015). Transforming to a Hyper-connected Society and Economy – Towards an "Industry 4.0." *Procedia Manufacturing*, 3, 417–424. https://doi.org/10.1016/j.promfg.2015.07.200

Benveniste, A. (2010). Loosely Time-Triggered Architectures for Cyber-Physical Systems. *2010 Design, Automation & Test in Europe Conference & Exhibition, Dresden*, 3–8. https://doi.org/doi: 10.1109/DATE.2010.5457246

Benveniste, A., Bouillard, A., & Caspi, P. (2010). A unifying view of loosely time-triggered architectures. *Proceedings of the Tenth ACM International Conference on Embedded Software - EMSOFT '10*, 189. https://doi.org/10.1145/1879021.1879047

Bhave, A., Krogh, B. H., Garlan, D., & Schmerl, B. (2011). View Consistency in Architectures for Cyber-Physical Systems. *2011 IEEE/ACM Second International Conference on Cyber-Physical Systems*, 151–160. https://doi.org/10.1109/ICCPS.2011.17

Biener, C., Eling, M., & Wirfs, J. H. (2014). Insurability of Cyber Risk 1. *The Geneva Association*, pp. 1–4. Retrieved from https://www.genevaassociation.org/media/891047/ga2014-if14-biener_elingwirfs.pdf

Blatter, J., & Haverland, M. (2012). *Designing Case Studies*. https://doi.org/10.1057/9781137016669

Bloem da Silveira Junior, L. A., Vasconcellos, E., Vasconcellos Guedes, L., Guedes, L. F. A., & Costa, R. M. (2018). Technology roadmapping: A methodological proposition to refine Delphi results. *Technological Forecasting and Social Change*, *126*, 194–206. https://doi.org/10.1016/J.TECHFORE.2017.08.011

Böhm, F., Menges, F., & Pernul, G. (2018). Graph-based visual analytics for cyber threat intelligence. *Cybersecurity*, *1*(1), 1–19. https://doi.org/10.1186/s42400-018-0017-4

Bouws, T., Kramer, F., Heemskerk, P., Van Os, M., Van Der Horst, T., Helmer, S., … De Heide, M. (2015). *Smart Industry: Dutch Industry Fit for the Future*. https://doi.org/527727

Boyes, H., Hallaq, B., Cunningham, J., & Watson, T. (2018). The industrial internet of things (IIoT): An analysis framework. *Computers in Industry*, *101*, 1–12. https://doi.org/10.1016/J.COMPIND.2018.04.015

Brass, I., Tanczer, L., Carr, M., Elsden, M., & Blackstock, J. (2018). Standardising a Moving Target: The Development and Evolution of IoT Security Standards. *Living in the Internet of Things: Cybersecurity of the IoT - 2018*, 24 (9 pp.)-24 (9 pp.). https://doi.org/10.1049/cp.2018.0024

Brass, Irina, Pothong, K., Tanczer, L., & Carr, M. (2019). *Standards, Governance and Policy. Cybersecurity of the Internet of Things (IoT): PETRAS Stream Report*. https://doi.org/10.13140/RG.2.2.15925.42729

Brettel, M., Fischer, F. G., Bendig, D., Weber, A. R., & Wolff, B. (2016). Enablers for Self-optimizing Production Systems in the Context of Industrie 4.0. *Procedia CIRP*, *41*, 93–98. https://doi.org/10.1016/j.procir.2015.12.065

Breza, M., Tomic, I., & McCann, J. (2018). Failures from the Environment, a Report on the First FAILSAFE workshop. *ACM SIGCOMM Computer Communication Review*, *48*(2), 40–45. https://doi.org/10.1145/3213232.3213238

Bryceson, K. P., & Slaughter, G. (2010). Alignment of performance metrics in a multi-enterprise agribusiness: achieving integrated autonomy? *International Journal of Productivity and Performance Management*, *59*(4), 325–350.

Carruthers, K. (2016). Internet of Things and Beyond: Cyber-Physical Systems - IEEE Internet of Things. *IEEE Internet of Things*. Retrieved from http://iot.ieee.org/newsletter/may-2016/internet-of-things-and-beyond-cyber-physical-systems.html

Charmaz, K. (2006). *Constructing grounded theory : a practical guide through qualitative analysis*. Sage Publications.

Córdova, F., Durán, C., Sepúlveda, J., Fernández, A., & Rojas, M. (2012). A proposal of logistic services innovation strategy for a mining company. *Journal of Technology Management & Innovation*, *7*(1), 175–185.

CRACS. (2018). Petras - Cyber Risk Assessment for Coupled Systems (CRACS). Retrieved February 20, 2020, from EPSRC website: https://petras-iot.org/project/cyber-risk-assessment-for-coupled-systems-cracs/

Craggs, B., & Rashid, A. (2017). Smart Cyber-Physical Systems: Beyond Usable Security to Security Ergonomics by Design. *2017 IEEE/ACM 3rd International Workshop on Software Engineering for Smart Cyber-Physical Systems (SEsCPS)*, 22–25. https://doi.org/10.1109/SEsCPS.2017.5

CVSS. (2019). Common Vulnerability Scoring System SIG. Retrieved December 26, 2017, from FIRST.org website: https://www.first.org/cvss/

David, M. (2005). *Science in Society*. New York: Palgrave Macmillan.

DCMS. (2017). *UK Digital Strategy 2017 - GOV.UK; Department for Culture, Media and Sport*. Retrieved from https://www.gov.uk/government/publications/uk-digital-strategy/uk-digital-strategy

de Reuver, M., Sørensen, C., & Basole, R. C. (2017). The digital platform: a research agenda. *Journal of Information Technology*, 33(2), 1–12. https://doi.org/10.1057/s41265-016-0033-3

De Roure, D., Page, K. R., Radanliev, P., & Van Kleek, M. (2019). Complex coupling in cyber-physical systems and the threats of fake data. *Living in the Internet of Things (IoT 2019), 2019 Page*, 11 (6 pp.). https://doi.org/10.1049/cp.2019.0136

De Roure, David, Hendler, J. A., James, D., Nurmikko-Fuller, T., Van Kleek, M., & Willcox, P. (2019). Towards a cyberphysicalweb science: A social machines perspective on pokémon go! *WebSci 2019 - Proceedings of the 11th ACM Conference on Web Science*, 65–69. https://doi.org/10.1145/3292522.3326043

De Roure, David, Hooper, C., Page, K., Tarte, S., & Willcox, P. (2015). Observing Social Machines Part 2. *Proceedings of the ACM Web Science Conference on ZZZ - WebSci '15*, 1–5. https://doi.org/10.1145/2786451.2786475

Dillon, T. S., Zhuge, H., Wu, C., Singh, J., & Chang, E. (2011). Web-of-things framework for cyber-physical systems. *Concurrency and Computation: Practice and Experience*, 23(9), 905–923. https://doi.org/10.1002/cpe.1629

DiMase, D., Collier, Z. A., Heffner, K., & Linkov, I. (2015). Systems engineering framework for cyber physical security and resilience. *Environment Systems and Decisions*, 35(2), 291–300. https://doi.org/10.1007/s10669-015-9540-y

Dombrowski, U., & Wagner, T. (2014). Mental Strain as Field of Action in the 4th Industrial Revolution. *Procedia CIRP*, 17, 100–105. https://doi.org/10.1016/j.procir.2014.01.077

Dotoli, M., Fanti *, M. P., Meloni, C., & Zhou, M. C. (2005). A multi-level approach for network design of integrated supply chains. *International Journal of Production Research*, 43(20), 4267–4287. https://doi.org/10.1080/00207540500142316

Easterby-Smith, M., Thorpe, R., & Lowe, A. (2002). *Management research : an introduction*. SAGE.

Eggenschwiler, J., Agrafiotis, I., & Nurse, J. R. (2016). Insider threat response and recovery strategies in financial services firms. *Computer Fraud & Security*, 2016(11), 12–19. https://doi.org/10.1016/S1361-3723(16)30091-4

Eisenhardt, K. M. (1989). Building Theories from Case Study Research. *The Academy of Management Review*, 14(4), 532. https://doi.org/10.2307/258557

Eriksson, P., & Kovalainen, A. (2008). *Qualitative methods in business research*. SAGE.

Evans, P. C., & Annunziata, M. (2012). *Industrial Internet: Pushing the Boundaries of Minds and Machines*. Retrieved from https://www.ge.com/docs/chapters/Industrial_Internet.pdf

FAIR. (2020). FAIR Risk Analytics Platform Management. Retrieved December 26, 2017, from FAIR-U Model website: https://www.fairinstitute.org/fair-u

Faller, C., & Feldmüller, D. (2015). Industry 4.0 Learning Factory for regional SMEs. *Procedia CIRP*,

Radanliev, Petar, David De Roure, Kevin Page, Jason R.C. Nurse, Rafael Mantilla Montalvo, Omar Santos, La'Treall Maddox, and Pete Burnap. "Cyber Risk at the Edge: Current and Future Trends on Cyber Risk Analytics and Artificial Intelligence in the Industrial Internet of Things and Industry 4.0 Supply Chains." *Cybersecurity, Springer Nature*, 2020. https://doi.org/10.1186/s42400-020-00052-8.

*32*, 88–91. https://doi.org/10.1016/j.procir.2015.02.117

Frohlich, M., & Westbrook, R. (2001). *Arcs of Integration: An International Study of Supply Chain Strategies*.

FTSE Russell. (2018). Industry Classification Benchmark (ICB) | FTSE Russell. Retrieved February 23, 2018, from FTSE International Limited and Frank Russell Company website: http://www.ftserussell.com/financial-data/industry-classification-benchmark-icb

G20. (2016). *G20 New Industrial Revolution Action Plan*. Retrieved from http://g20chn.org/English/Documents/Current/201609/P020160908738867573193.pdf

Gershenfeld, N. A. (1999). *When things start to think*. Retrieved from https://books.google.com/books?hl=en&lr=&id=J8GLAwAAQBAJ&oi=fnd&pg=PP2&dq=When+Things+Start+to+Think&ots=8HHfEEuYYh&sig=vSgqQS_0PtX0cH_E_d0uDVTYlCI#v=onepage&q=When Things Start to Think&f=false

Ghirardello, K., Maple, C., Ng, D., & Kearney, P. (2018). Cyber security of smart homes: development of a reference architecture for attack surface analysis. *Living in the Internet of Things: Cybersecurity of the IoT - 2018*, 45 (10 pp.)-45 (10 pp.). https://doi.org/10.1049/cp.2018.0045

Giordano, A., Spezzano, G., & Vinci, A. (2016). *A Smart Platform for Large-Scale Cyber-Physical Systems*. https://doi.org/10.1007/978-3-319-26869-9_6

Glaser, B. G., & Strauss, A. L. (1967). *The discovery of grounded theory : strategies for qualitative research*. Abingdon, Oxford: Routledge.

Gordon, L. A., & Loeb, M. P. (2002). The economics of information security investment. *ACM Transactions on Information and System Security*, *5*(4), 438–457. https://doi.org/10.1145/581271.581274

Goulding, C. (2002). *Grounded theory : a practical guide for management, business and market researchers*. SAGE.

GTAI. (2014). *Industrie 4.0 Smart Manufacturing for the Future*. Retrieved from https://www.gtai.de/GTAI/Content/EN/Invest/_SharedDocs/Downloads/GTAI/Brochures/Industries/industrie4.0-smart-manufacturing-for-the-future-en.pdf

Gubbi, J., Buyya, R., Marusic, S., & Palaniswami, M. (2013). Internet of Things (IoT): A vision, architectural elements, and future directions. *Future Generation Computer Systems*, *29*(7), 1645–1660. https://doi.org/10.1016/j.future.2013.01.010

Gummesson, E. (2000). *Qualitative methods in management research*. Sage.

Hahn, A., Ashok, A., Sridhar, S., & Govindarasu, M. (2013). Cyber-Physical Security Testbeds: Architecture, Application, and Evaluation for Smart Grid. *IEEE Transactions on Smart Grid*, *4*(2), 847–855. https://doi.org/10.1109/TSG.2012.2226919

Hermann, M., Pentek, T., & Otto, B. (2016). Design Principles for Industrie 4.0 Scenarios. *2016 49th Hawaii International Conference on System Sciences (HICSS)*, 3928–3937. https://doi.org/10.1109/HICSS.2016.488

Hussain, F. (2017). Internet of Everything. In *Internet of Things: Building Blocks and Business Models: SpringerBriefs in Electrical and Computer Engineering* (pp. 1–11). https://doi.org/10.1007/978-3-319-55405-1_1

IAM. (2018). Petras - Impact Assessment Model for the IoT (IAM). Retrieved February 20, 2020, from EPSRC website: https://petras-iot.org/project/impact-assessment-model-for-the-iot-iam/

IIC. (2016). *The Industrial Internet of Things, Volume B01: Business Strategy and Innovation Framework; Industrial Internet Consortium*. https://doi.org/IIC:PUB:B01:V1.0:PB:20161115

IIC. (2017). *The Industrial Internet of Things Volume G5: Connectivity Framework; Industrial Internet Consortium*. Retrieved from http://www.iiconsortium.org/pdf/IIC_PUB_G5_V1.0_PB_20170228.pdf

Radanliev, Petar, David De Roure, Kevin Page, Jason R.C. Nurse, Rafael Mantilla Montalvo, Omar Santos, La'Treall Maddox, and Pete Burnap. "Cyber Risk at the Edge: Current and Future Trends on Cyber Risk Analytics and Artificial Intelligence in the Industrial Internet of Things and Industry 4.0 Supply Chains." *Cybersecurity, Springer Nature*, 2020. https://doi.org/10.1186/s42400-020-00052-8.

Industrie 4.0. (2017). Plattform Industrie 4.0 - Testbeds. Retrieved May 13, 2017, from http://www.plattform-i40.de/I40/Navigation/EN/InPractice/Testbeds/testbeds.html

IVI. (2017). *Industrial Value Chain Reference Architecture; Industrial Value Chain Initiative*. Retrieved from https://iv-i.org/en/docs/Industrial_Value_Chain_Reference_Architecture_170424.pdf

Jayaram, J., & Tan, K.-C. (2010). Supply chain integration with third-party logistics providers. *International Journal of Production Economics*, *125*(2), 262–271.

Jazdi, N. (2014). Cyber physical systems in the context of Industry 4.0. *2014 IEEE International Conference on Automation, Quality and Testing, Robotics*, 1–4. https://doi.org/10.1109/AQTR.2014.6857843

Jensen, J. C., Chang, D. H., & Lee, E. A. (2011). A model-based design methodology for cyber-physical systems. *2011 7th International Wireless Communications and Mobile Computing Conference*, 1666–1671. https://doi.org/10.1109/IWCMC.2011.5982785

John, P. (2017). *High Value Manufacturing Catapult*. Retrieved from https://ec.europa.eu/growth/tools-databases/regional-innovation-monitor/sites/default/files/report/High Value Manufacturing Catapult_1.pdf

Kambatla, K., Kollias, G., Kumar, V., & Grama, A. (2014). Trends in big data analytics. *J. Parallel Distrib. Comput*, *74*, 2561–2573. https://doi.org/10.1016/j.jpdc.2014.01.003

Kang, W., Kapitanova, K., & Son, S. H. (2012). RDDS: A Real-Time Data Distribution Service for Cyber-Physical Systems. *IEEE Transactions on Industrial Informatics*, *8*(2), 393–405. https://doi.org/10.1109/TII.2012.2183878

Kaplan, R. S., & Norton, D. P. (1996). *Using the balanced scorecard as a strategic management system*. Harvard business review Boston.

Kirkpatrick, K. (2013). Software-defined networking. *Communications of the ACM*, *56*(9), 16. https://doi.org/10.1145/2500468.2500473

Koch, R., & Rodosek, G. (2016). *Proceedings of the 15th European Conference on Cyber Warfare and Security : ECCWS 2016 : hosted by Universität der Bundeswehr, Munich, Germany 7-8 July 2016*. Retrieved from https://books.google.co.uk/books?hl=en&lr=&id=ijaeDAAAQBAJ&oi=fnd&pg=PA145&dq=economic+impact+of+cyber+risk&ots=50mTo8TVSV&sig=sD4V76yG5tG6IZIglmnGz3L1qqw&redir_esc=y#v=onepage&q=economic impact of cyber risk&f=false

Kolberg, D., & Zühlke, D. (2015). Lean Automation enabled by Industry 4.0 Technologies. *IFAC-PapersOnLine*, *48*(3), 1870–1875. https://doi.org/10.1016/j.ifacol.2015.06.359

La, H. J., & Kim, S. D. (2010). A Service-Based Approach to Designing Cyber Physical Systems. *2010 IEEE/ACIS 9th International Conference on Computer and Information Science*, 895–900. https://doi.org/10.1109/ICIS.2010.73

Lee, B., Cooper, R., Hands, D., & Coulton, P. (2019a). Design Drivers: A critical enabler to meditate value over the NPD process within Internet of Things. *4d Conference Proceedings: Meanings of Design in the Next Era. Osaka : DML (Design Management Lab), Ritsumeikan University*, 96–107. Osaka.

Lee, B., Cooper, R., Hands, D., & Coulton, P. (2019b). Value creation for IoT: Challenges and opportunities within the design and development process. *Living in the Internet of Things (IoT 2019). IET, Living in the Internet of Things 2019, London, United Kingdom*, 1–8. Retrieved from https://doi.org/10.1049/cp.2019.0127

Lee, J., Bagheri, B., & Kao, H.-A. (2015). A Cyber-Physical Systems architecture for Industry 4.0-based manufacturing systems. In *Manufacturing Letters* (Vol. 3). https://doi.org/10.1016/j.mfglet.2014.12.001

Pre-print – before proofread by journal print production team.
Reference:

Radanliev, Petar, David De Roure, Kevin Page, Jason R.C. Nurse, Rafael Mantilla Montalvo, Omar Santos, La'Treall Maddox, and Pete Burnap. "Cyber Risk at the Edge: Current and Future Trends on Cyber Risk Analytics and Artificial Intelligence in the Industrial Internet of Things and Industry 4.0 Supply Chains." *Cybersecurity, Springer Nature*, 2020. https://doi.org/10.1186/s42400-020-00052-8.

Lee, J., Kao, H.-A., & Yang, S. (2014). Service Innovation and Smart Analytics for Industry 4.0 and Big Data Environment. *Procedia CIRP*, *16*, 3–8. https://doi.org/10.1016/j.procir.2014.02.001

Leitão, P., Colombo, A. W., & Karnouskos, S. (2016). Industrial automation based on cyber-physical systems technologies: Prototype implementations and challenges. *Computers in Industry*, *81*, 11–25. https://doi.org/10.1016/j.compind.2015.08.004

Leng, K., & Chen, X. (2012). A genetic algorithm approach for TOC-based supply chain coordination. *Applied Mathematics and Information Sciences*, *6*(3), 767–774.

Leonard, T. C. (2008). Richard H. Thaler, Cass R. Sunstein, Nudge: Improving decisions about health, wealth, and happiness. *Constitutional Political Economy*, *19*(4), 356–360. https://doi.org/10.1007/s10602-008-9056-2

Lewis, D., & Brigder, D. (2004). Market Researchers make Increasing use of Brain Imaging. *Advances in Clinical Neuroscience and Rehabilitation*, *5*(3), 36–37. Retrieved from http://www.acnr.co.uk/pdfs/volume5issue3/v5i3specfeat.pdf

Li, L. (2017). China's manufacturing locus in 2025: With a comparison of "Made-in-China 2025" and "Industry 4.0." *Technological Forecasting and Social Change*. https://doi.org/10.1016/J.TECHFORE.2017.05.028

Li, W., Liu, K., Belitski, M., Ghobadian, A., & O'Regan, N. (2016). e-Leadership through strategic alignment: an empirical study of small- and medium-sized enterprises in the digital age. *Journal of Information Technology*, *31*(2), 185–206. https://doi.org/10.1057/jit.2016.10

Longstaff, T. A., & Haimes, Y. Y. (2002). A holistic roadmap for survivable infrastructure systems. *IEEE Transactions on Systems, Man, and Cybernetics - Part A: Systems and Humans*, *32*(2), 260–268. https://doi.org/10.1109/TSMCA.2002.1021113

Lu, H.-P., & Weng, C.-I. (2018). Smart manufacturing technology, market maturity analysis and technology roadmap in the computer and electronic product manufacturing industry. *Technological Forecasting and Social Change*. https://doi.org/10.1016/j.techfore.2018.03.005

Madaan, A., Nurse, J., de Roure, D., O'Hara, K., Hall, W., & Creese, S. (2018). A Storm in an IoT Cup: The Emergence of Cyber-Physical Social Machines. *SSRN Electronic Journal*. https://doi.org/10.2139/ssrn.3250383

Madakam, S., Ramaswamy, R., & Tripathi, S. (2015). Internet of Things (IoT): A Literature Review. *Journal of Computer and Communications*, *3*(3), 164–173. https://doi.org/10.4236/jcc.2015.35021

Manthou, V., Vlachopoulou, M., & Folinas, D. (2004). Virtual e-Chain (VeC) model for supply chain collaboration. *International Journal of Production Economics*, *87*(3), 241–250.

Maple, C., Bradbury, M., Le, A. T., & Ghirardello, K. (2019). A Connected and Autonomous Vehicle Reference Architecture for Attack Surface Analysis. *Applied Sciences*, *9*(23), 5101. https://doi.org/10.3390/app9235101

Marwedel, P., & Engel, M. (2016). *Cyber-Physical Systems: Opportunities, Challenges and (Some) Solutions*. https://doi.org/10.1007/978-3-319-26869-9_1

MEICA. (2015). *Industria Conectada 4.0: La transformación digital de la industria española Dossier de prensa; Ministry of Economy Industry and Competitiveness Accessibility*. Retrieved from http://www.lamoncloa.gob.es/serviciosdeprensa/notasprensa/Documents/081015 Dossier prensa Industria 4 0.pdf

Melnyk, S. A., Narasimhan, R., & DeCampos, H. A. (2014). Supply chain design: issues, challenges, frameworks and solutions. *International Journal of Production Research*, *52*(7), 1887–1896. https://doi.org/10.1080/00207543.2013.787175

Mentzer, J. T., DeWitt, W., Keebler, J. S., Min, S., Nix, N. W., Smith, C. D., & Zacharia, Z. G. (2001). Defining supply chain management. In *Journal of Business logistics* (Vol. 22). Wiley Online

Library.

Metallo, C., Agrifoglio, R., Schiavone, F., & Mueller, J. (2018). Understanding business model in the Internet of Things industry. *Technological Forecasting and Social Change*. https://doi.org/10.1016/J.TECHFORE.2018.01.020

METI. (2015). *NRS, New Robot Strategy - Vision Strategy and Action Plan; Ministry of Economy Trade and Industry of Japan*. Retrieved from http://www.meti.go.jp/english/press/2015/pdf/0123_01b.pdf

METIJ. (2015). *RRI, Robot Revolution Initiative - Summary of Japan's Robot Strategy - It's vision, strategy and action plan; Ministry of Economy, Trade and Industry of Japan*. Retrieved from http://www.meti.go.jp/english/press/2015/pdf/0123_01c.pdf

Miles, M. B., Huberman, A. M., & Saldaña, J. (1983). *Qualitative data analysis : a methods sourcebook*.

MIUR. (2014). Italian Technology Cluster: Intelligent Factories; Ministry of Education Universities and Research. Retrieved May 9, 2017, from Cluster Tecnologico Nazionale Fabbrica Intelligente | Imprese, università, organismi di ricerca, associazioni e enti territoriali: insieme per la crescita del Manifatturiero website: http://www.fabbricaintelligente.it/en/

Müller, J. M., Buliga, O., & Voigt, K.-I. (2018). Fortune favors the prepared: How SMEs approach business model innovations in Industry 4.0. *Technological Forecasting and Social Change*. https://doi.org/10.1016/J.TECHFORE.2017.12.019

Nicolescu, R., Huth, M., Radanliev, P., & De Roure, D. (2018a). Mapping the values of IoT. *Journal of Information Technology*, *33*(4), 345–360. https://doi.org/10.1057/s41265-018-0054-1

Nicolescu, R., Huth, M., Radanliev, P., & De Roure, D. (2018b). *State of The Art in IoT - Beyond Economic Value*. Retrieved from https://iotuk.org.uk/wp-content/uploads/2018/08/State-of-the-Art-in-IoT-–-Beyond-Economic-Value2.pdf

NIF. (2016). *New Industrial France: Building France's industrial future - updated text from the 2013 version*. Retrieved from https://www.economie.gouv.fr/files/files/PDF/web-dp-indus-ang.pdf

Niggemann, O., Biswas, G., Kinnebrew, J. S., Khorasgani, H., Volgmann, S., & Bunte, A. (2015). Data-Driven Monitoring of Cyber-Physical Systems Leveraging on Big Data and the Internet-of-Things for Diagnosis and Control. *International Workshop on the Principles of Diagnosis (DX)*, 185–192. Retrieved from http://ceur-ws.org/Vol-1507/dx15paper24.pdf

Nikulin, C., Graziosi, S., Cascini, G., Araneda, A., & Minutolo, M. (2013). An algorithm for supply chain integration based on OTSM-TRIZ. *Procedia-Social and Behavioral Sciences*, *75*, 383–396.

Nurse, J., Creese, S., & De Roure, D. (2017). Security Risk Assessment in Internet of Things Systems. *IT Professional*, *19*(5), 20–26. https://doi.org/10.1109/MITP.2017.3680959

Nurse, J. R., Radanliev, P., Creese, S., & De Roure, D. (2018). Realities of Risk: 'If you can't understand it, you can't properly assess it!': The reality of assessing security risks in Internet of Things systems. *Living in the Internet of Things: Cybersecurity of the IoT - 2018*, 1–9. https://doi.org/10.1049/cp.2018.0001

Okutan, A., Werner, G., Yang, S. J., & McConky, K. (2018). Forecasting cyberattacks with incomplete, imbalanced, and insignificant data. *Cybersecurity*, *1*(1), 1–16. https://doi.org/10.1186/s42400-018-0016-5

Okutan, A., & Yang, S. J. (2019). ASSERT: attack synthesis and separation with entropy redistribution towards predictive cyber defense. *Cybersecurity*, *2*(1), 1–18. https://doi.org/10.1186/s42400-019-0032-0

Ouyang, J., Lin, S., Jiang, S., Hou, Z., Wang, Y., Wang, Y., … Hou, Zhenyu; Wang, Yong; Wang, Y. (2014). SDF: software-defined flash for web-scale internet storage systems. *Proceedings of the 19th International Conference on Architectural Support for Programming Languages and Operating Systems - ASPLOS '14*, *42*(1), 471–484. https://doi.org/10.1145/2541940.2541959

Paltridge, B. (2017). Peer Review in Academic Settings. In *The Discourse of Peer Review* (pp. 1–29). https://doi.org/10.1057/978-1-137-48736-0_1

Pan, M., Sikorski, J., Kastner, C. A., Akroyd, J., Mosbach, S., Lau, R., & Kraft, M. (2015). Applying Industry 4.0 to the Jurong Island Eco-industrial Park. *Energy Procedia*, *75*, 1536–1541. https://doi.org/10.1016/j.egypro.2015.07.313

Perez-Franco, R. (2016). *Rethinking your supply chain strategy: a brief guide*.

PETRAS. (2020). Impact of Cyber Risk at the Edge: Cyber Risk Analytics and Artificial Intelligence (CRatE). Retrieved February 17, 2020, from https://petras-iot.org/project/impact-of-cyber-risk-at-the-edge-cyber-risk-analytics-and-artificial-intelligence-crate/

Petrolo, R., Loscri, V., & Mitton, N. (2016). *Cyber-Physical Objects as Key Elements for a Smart Cyber-City*. https://doi.org/10.1007/978-3-319-26869-9_2

Posada, J., Toro, C., Barandiaran, I., Oyarzun, D., Stricker, D., de Amicis, R., … Vallarino, I. (2015). Visual Computing as a Key Enabling Technology for Industrie 4.0 and Industrial Internet. *IEEE Computer Graphics and Applications*, *35*(2), 26–40. https://doi.org/10.1109/MCG.2015.45

Prajogo, D., & Olhager, J. (2012). Supply chain integration and performance: The effects of long-term relationships, information technology and sharing, and logistics integration. *International Journal of Production Economics*, *135*(1), 514–522.

Pramatari, K., Evgeniou, T., & Doukidis, G. (2009). Implementation of collaborative e-supply-chain initiatives: an initial challenging and final success case from grocery retailing. *Journal of Information Technology*, *24*(3), 269–281. https://doi.org/10.1057/jit.2008.11

Qu, T., Huang, G. Q., Cung, V.-D., & Mangione, F. (2010). Optimal configuration of assembly supply chains using analytical target cascading. *International Journal of Production Research*, *48*(23), 6883–6907. https://doi.org/10.1080/00207540903307631

Radanliev, P., De Roure, D., Nicolescu, R., & Huth, M. (2019). A reference architecture for integrating the Industrial Internet of Things in the Industry 4.0. In *University of Oxford combined working papers and project reports prepared for the PETRAS National Centre of Excellence and the Cisco Research Centre*. https://doi.org/10.13140/RG.2.2.26854.47686

Radanliev, P, Nicolescu, R., De Roure, D., & Huth, M. (2019). *Harnessing Economic Value from the Internet of Things*. London.

Radanliev, P, Roure, D. De, Nurse, J., & Nicolescu, R. (2019). Cyber risk impact assessment–discussion on assessing the risk from the IoT to the digital economy. *University of Oxford Combined Working Papers and Project Reports Prepared for the PETRAS National Centre of Excellence and the Cisco Research Centre*.

Radanliev, Petar. (2014). *A conceptual framework for supply chain systems architecture and integration design based on practice and theory in the North Wales slate mining industry* (British Library). https://doi.org/ISNI: 0000 0004 5352 6866

Radanliev, Petar. (2015a). Architectures for Green-Field Supply Chain Integration. *Journal of Supply Chain and Operations Management*, *13*(2). https://doi.org/10.20944/preprints201904.0144.v1

Radanliev, Petar. (2015b). Engineering Design Methodology for Green-Field Supply Chain Architectures Taxonomic Scheme. *Journal of Operations and Supply Chain Management*, *8*(2), 52–66. https://doi.org/10.12660/joscmv8n2p52-66

Radanliev, Petar. (2015c). Green-field Architecture for Sustainable Supply Chain Strategy Formulation. *International Journal of Supply Chain Management*, *4*(2), 62–67. https://doi.org/10.20944/preprints201904.0116.v1

Radanliev, Petar. (2016). Supply Chain Systems Architecture and Engineering Design: Green-field Supply Chain Integration. *Operations and Supply Chain Management: An International Journal*, *9*(1). https://doi.org/10.20944/preprints201904.0122.v1

Radanliev, Petar. (2019a). CYBER RISK IMPACT ASSESSMENT. In *University of Oxford combined working papers and project reports prepared for the PETRAS National Centre of Excellence and the Cisco Research Centre*. Oxford, University of Oxford combined working papers and project reports prepared for the PETRAS National Centre of Excellence and the Cisco Research Centre.

Radanliev, Petar. (2019b). Cyber Risk Management for the Internet of Things. In *University of Oxford combined working papers and project reports prepared for the PETRAS National Centre of Excellence and the Cisco Research Centre.* https://doi.org/10.13140/RG.2.2.34482.86722

Radanliev, Petar. (2019c). Digital Supply Chains for Industry 4.0 Taxonomy of Approaches. *University of Oxford Combined Working Papers and P*, (April). https://doi.org/10.20944/preprints201904.0160.v1

Radanliev, Petar, Charles De Roure, D., Nurse, J. R. C., Burnap, P., & Montalvo, R. M. (2019). Methodology for designing decision support supply chain systems for visualising and mitigating cyber risk from IoT technologies. In *University of Oxford combined working papers and project reports prepared for the PETRAS National Centre of Excellence and the Cisco Research Centre.* https://doi.org/10.13140/RG.2.2.32975.53921

Radanliev, Petar, De Roure, D. C., Nurse, J. R. C., Montalvo, R. M., & Burnap, P. (2019a). The Industrial Internet-of-Things in the Industry 4.0 supply chains of small and medium sized enterprises. In *University of Oxford combined working papers and project reports prepared for the PETRAS National Centre of Excellence and the Cisco Research Centre.* https://doi.org/10.13140/RG.2.2.14140.49283

Radanliev, Petar, De Roure, D. C., Nurse, J. R. C., Montalvo, R. M., Burnap, P., Roure, D. C. De, … Montalvo, R. M. (2019). Design principles for cyber risk impact assessment from Internet of Things (IoT). In *University of Oxford combined working papers and project reports prepared for the PETRAS National Centre of Excellence and the Cisco Research Centre.* https://doi.org/10.13140/RG.2.2.33014.86083

Radanliev, Petar, De Roure, D., Cannady, S., Mantilla Montalvo, R., Nicolescu, R., & Huth, M. (2018). Economic impact of IoT cyber risk - analysing past and present to predict the future developments in IoT risk analysis and IoT cyber insurance. *Living in the Internet of Things: Cybersecurity of the IoT - 2018*, (CP740), 3 (9 pp.). https://doi.org/10.1049/cp.2018.0003

Radanliev, Petar, De Roure, D., Cannady, S., Montalvo, R. M., Nicolescu, R., & Huth, M. (2019). Analysing IoT cyber risk for estimating IoT cyber insurance. *University of Oxford Combined Working Papers and Project Reports Prepared for the PETRAS National Centre of Excellence and the Cisco Research Centre.* https://doi.org/10.13140/RG.2.2.25006.36167

Radanliev, Petar, De Roure, D., Maple, C., Nurse, J. R. ., Nicolescu, R., & Ani, U. (2019). Cyber Risk in IoT Systems. In *University of Oxford combined working papers and project reports prepared for the PETRAS National Centre of Excellence and the Cisco Research Centre.* https://doi.org/10.13140/RG.2.2.29652.86404

Radanliev, Petar, De Roure, D., Nicolescu, R., Huth, M., Montalvo, R. M., Cannady, S., & Burnap, P. (2018). Future developments in cyber risk assessment for the internet of things. *Computers in Industry*, *102*, 14–22. https://doi.org/10.1016/J.COMPIND.2018.08.002

Radanliev, Petar, De Roure, D., Nurse, J. R. ., Nicolescu, R., Huth, M., Cannady, S., & Mantilla Montalvo, R. (2018). Integration of Cyber Security Frameworks, Models and Approaches for Building Design Principles for the Internet-of-things in Industry 4.0. *Living in the Internet of Things: Cybersecurity of the IoT*, 41 (6 pp.). https://doi.org/10.1049/cp.2018.0041

Radanliev, Petar, De Roure, D., Nurse, J. R. ., Nicolescu, R., Huth, M., Cannady, S., & Mantilla Montalvo, R. (2019a). New developments in Cyber Physical Systems, the Internet of Things and the Digital Economy – future developments in the Industrial Internet of Things and Industry 4.0.

Radanliev, Petar, David De Roure, Kevin Page, Jason R.C. Nurse, Rafael Mantilla Montalvo, Omar Santos, La'Treall Maddox, and Pete Burnap. "Cyber Risk at the Edge: Current and Future Trends on Cyber Risk Analytics and Artificial Intelligence in the Industrial Internet of Things and Industry 4.0 Supply Chains." *Cybersecurity, Springer Nature*, 2020. https://doi.org/10.1186/s42400-020-00052-8.

In *University of Oxford combined working papers and project reports prepared for the PETRAS National Centre of Excellence and the Cisco Research Centre*. https://doi.org/10.13140/RG.2.2.14133.93921

Radanliev, Petar, De Roure, D., Nurse, J. R., Burnap, P., Anthi, E., Ani, U., … Mantilla Montalvo, R. (2019b). Cyber risk from IoT technologies in the supply chain-discussion on supply chains decision support system for the digital economy. In *University of Oxford combined working papers and project reports prepared for the PETRAS National Centre of Excellence and the Cisco Research Centre*. https://doi.org/10.13140/RG.2.2.17286.22080

Radanliev, Petar, De Roure, D., Nurse, J. R. C., Montalvo, R. M., & Burnap, P. (2019b). Standardisation of cyber risk impact assessment for the Internet of Things (IoT). In *University of Oxford combined working papers and project reports prepared for the PETRAS National Centre of Excellence and the Cisco Research Centre*. https://doi.org/10.13140/RG.2.2.27903.05280

Radanliev, Petar, De Roure, D., Nurse, J. R. C., Nicolescu, R., Huth, M., Cannady, S., & Montalvo, R. M. (2019c). Cyber Security Framework for the Internet-of-Things in Industry 4.0. In *University of Oxford combined working papers and project reports prepared for the PETRAS National Centre of Excellence and the Cisco Research Centre*. https://doi.org/10.13140/RG.2.2.32955.87845

Radanliev, Petar, DeRoure, D., Nurse, J. R. C., Burnap, P., Anthi, E., Ani, U., … Montalvo, R. M. (2019). Definition of Cyber Strategy Transformation Roadmap for Standardisation of IoT Risk Impact Assessment with a Goal-Oriented Approach and the Internet of Things Micro Mart. In *University of Oxford combined working papers and project reports prepared for the PETRAS National Centre of Excellence and the Cisco Research Centre*. https://doi.org/10.13140/RG.2.2.12462.77124

Radanliev, Petar, Roure, D. C. De, R.C. Nurse, J., Montalvo, R. M., Cannady, S., Santos, O., … Maple, C. (2020). Future developments in standardisation of cyber risk in the Internet of Things (IoT). *SN Applied Sciences*, (2: 169), 1–16. https://doi.org/10.1007/s42452-019-1931-0

Radanliev, Petar, Roure, D. De, Nurse, J. R. C. C., Nicolescu, R., Huth, M., Cannady, S., … Montalvo, R. M. (2019). Cyber Risk impact Assessment - Assessing the Risk from the IoT to the Digital Economy. In *University of Oxford combined working papers and project reports prepared for the PETRAS National Centre of Excellence and the Cisco Research Centre*. https://doi.org/10.13140/RG.2.2.11145.49768

Rajkumar, R., Lee, I., Sha, L., & Stankovic, J. (2010). Cyber-Physical Systems: The Next Computing Revolution. *Proceedings of the 47th Design Automation Conference on - DAC '10*, 731. https://doi.org/10.1145/1837274.1837461

Ribeiro, L., Barata, J., & Ferreira, J. (2010). An agent-based interaction-oriented shop floor to support emergent diagnosis. *2010 8th IEEE International Conference on Industrial Informatics*, 189–194. https://doi.org/10.1109/INDIN.2010.5549436

Ringert, J. O., Rumpe, B., & Wortmann, A. (2015). *Architecture and Behavior Modeling of Cyber-Physical Systems with MontiArcAutomaton*. Retrieved from http://arxiv.org/abs/1509.04505

Rodewald, G., & Gus. (2005). Aligning information security investments with a firm's risk tolerance. *Proceedings of the 2nd Annual Conference on Information Security Curriculum Development - InfoSecCD '05*, 139. https://doi.org/10.1145/1107622.1107654

Rosenzweig, E. D., Roth, A. V, & Dean, J. W. (2003). The influence of an integration strategy on competitive capabilities and business performance: an exploratory study of consumer products manufacturers. *Journal of Operations Management*, 21(4), 437–456.

Roumani, M. A., Fung, C. C., Rai, S., & Xie, H. (2016). Value Analysis of Cyber Security Based on Attack Types. *ITMSOC Transactions on Innovation & Business Engineering*, 01, 34–39.

Retrieved from http://www.itmsoc.org

Ruan, K. (2017). Introducing cybernomics: A unifying economic framework for measuring cyber risk. *Computers & Security*, *65*, 77–89. https://doi.org/10.1016/j.cose.2016.10.009

Rutter, T. (2015). The rise of nudge – the unit helping politicians to fathom human behavior. *The Guardian*, *7*(23), 2015. Retrieved from https://www.theguardian.com/public-leaders-network/2015/jul/23/rise-nudge-unit-politicians-human-behaviour

Safa, N. S., Maple, C., Watson, T., & Von Solms, R. (2018). Motivation and opportunity based model to reduce information security insider threats in organisations. *Journal of Information Security and Applications*, *40*, 247–257. https://doi.org/10.1016/J.JISA.2017.11.001

Sakka, O., Millet, P.-A., & Botta-Genoulaz, V. (2011). An ontological approach for strategic alignment: a supply chain operations reference case study. *International Journal of Computer Integrated Manufacturing*, *24*(11), 1022–1037.

Sangiovanni-Vincentelli, A., Damm, W., & Passerone, R. (2012). Taming Dr. Frankenstein: Contract-Based Design for Cyber-Physical Systems * g. *European Journal of Control*, *18*, 217–238. https://doi.org/10.3166/EJC.18.217–238

Schnetzler, M. J., Sennheiser, A., & Schönsleben, P. (2007). A decomposition-based approach for the development of a supply chain strategy. *International Journal of Production Economics*, *105*(1), 21–42. https://doi.org/10.1016/j.ijpe.2006.02.004

SCPRC. (2017). Made in China 2025; The State Council People Republic of China. Retrieved May 10, 2017, from www.english.gov.cn website: http://english.gov.cn/2016special/madeinchina2025/

Shackelford, S. J. (2016). Protecting Intellectual Property and Privacy in the Digital Age: The Use of National Cybersecurity Strategies to Mitigate Cyber Risk. *Chapman Law Review*, *19*, 412–445. Retrieved from http://heinonline.org/HOL/Page?handle=hein.journals/chlr19&id=469&div=26&collection=journals

Shadbolt, N., O'Hara, K., De Roure, D., & Hall, W. (2019). *The Theory and Practice of Social Machines*. In *Lecture Notes in Social Networks*. https://doi.org/10.1007/978-3-030-10889-2

Shafiq, S. I., Sanin, C., Szczerbicki, E., & Toro, C. (2015). Virtual Engineering Object / Virtual Engineering Process: A specialized form of Cyber Physical System for Industrie 4.0. *Procedia Computer Science*, *60*, 1146–1155. https://doi.org/10.1016/j.procs.2015.08.166

Shaw, D. R., Snowdon, B., Holland, C. P., Kawalek, P., & Warboys, B. (2004). The viable systems model applied to a smart network: the case of the UK electricity market. *Journal of Information Technology*, *19*(4), 270–280. https://doi.org/10.1057/palgrave.jit.2000028

Shi, J., Wan, J., Yan, H., & Suo, H. (2011). A survey of Cyber-Physical Systems. *2011 International Conference on Wireless Communications and Signal Processing (WCSP)*, 1–6. https://doi.org/10.1109/WCSP.2011.6096958

Siemens. (2017). *Made Smarter review 2017*. Retrieved from https://assets.publishing.service.gov.uk/government/uploads/system/uploads/attachment_data/file/655570/20171027_MadeSmarter_FINAL_DIGITAL.pdf

Sirris and Agoria. (2017). Made Different: Factory of the Future 4.0. Retrieved May 9, 2017, from http://www.madedifferent.be/en/what-factory-future-40

Sokolov, B., & Ivanov, D. (2015). Integrated scheduling of material flows and information services in industry 4.0 supply networks. *IFAC-PapersOnLine*, *48*(3), 1533–1538. https://doi.org/10.1016/j.ifacol.2015.06.304

Stock, T., & Seliger, G. (2016). Opportunities of Sustainable Manufacturing in Industry 4.0. *Procedia CIRP*, *40*, 536–541. https://doi.org/10.1016/j.procir.2016.01.129

Stojmenovic, I. (2014). Machine-to-Machine Communications With In-Network Data Aggregation,

Processing, and Actuation for Large-Scale Cyber-Physical Systems. *IEEE Internet of Things Journal*, *1*(2), 122–128. https://doi.org/10.1109/JIOT.2014.2311693

Strader, T. J., Lin, F.-R., & Shaw, M. J. (1999). Business-to-business electronic commerce and convergent assembly supply chain management. *Journal of Information Technology*, *14*(4), 361–373. https://doi.org/10.1080/026839699344476

Sukati, I., Hamid, A. B., Baharun, R., & Yusoff, R. M. (2012). The study of supply chain management strategy and practices on supply chain performance. *Procedia-Social and Behavioral Sciences*, *40*, 225–233.

Tan, Y., Goddard, S., & Pérez, L. C. (2008). A Prototype Architecture for Cyber-Physical Systems. *ACM SIGBED Review - Special Issue on the RTSS Forum on Deeply Embedded Real-Time Computing*, *5*(1). Retrieved from http://delivery.acm.org/10.1145/1370000/1366309/p26-tan.pdf?ip=129.67.116.155&id=1366309&acc=ACTIVE SERVICE&key=BF07A2EE685417C5.F2FAECDC86A918EB.4D4702B0C3E38B35.4D4702B0C3E38B35&CFID=922793771&CFTOKEN=47199625&__acm__=1492383641_ca27b2c456d59140

Tanczer, L. M., Steenmans, I., Elsden, M., Blackstock, J., & Carr, M. (2018). Emerging risks in the IoT ecosystem: Who's afraid of the big bad smart fridge? *Living in the Internet of Things: Cybersecurity of the IoT*, 33 (9 pp.). https://doi.org/10.1049/cp.2018.0033

Taylor, P., Allpress, S., Carr, M., Lupu, E., Norton, J., Smith, L., Blackstock, J., Boyes, H., Hudson-Smith, A., Brass, I., Chizari, H., Cooper, R., Coulton, P., Craggs, B., Davies, N., De Roure, D., Elsden, M., Huth, M., Lindley, J., Maple, C., Mittelstadt, B., Nicolescu, R., Nurse, J., Procter, R., Radanliev, P., Rashid, A., Sgandurra, D., Skatova, A., Taddeo, M., Tanczer, L., Vieira-Steiner, R., … R.J., Westbury, P. S. (2018). *Internet of Things realising the potential of a trusted smart world*. London.

Thramboulidis, K. (2015). A cyber–physical system-based approach for industrial automation systems. *Computers in Industry*, *72*, 92–102. https://doi.org/10.1016/j.compind.2015.04.006

Toro, C., Barandiaran, I., & Posada, J. (2015). A Perspective on Knowledge Based and Intelligent Systems Implementation in Industrie 4.0. *Procedia Computer Science*, *60*, 362–370. https://doi.org/10.1016/j.procs.2015.08.143

Van der Vaart, T., & van Donk, D. P. (2008). A critical review of survey-based research in supply chain integration. *International Journal of Production Economics*, *111*(1), 42–55.

Van Kleek, M., Binns, R., Zhao, J., Slack, A., Lee, S., Ottewell, D., & Shadbolt, N. (2018). X-Ray Refine. *Proceedings of the 2018 CHI Conference on Human Factors in Computing Systems - CHI '18*, 1–13. https://doi.org/10.1145/3173574.3173967

Vickery, S. K., Jayaram, J., Droge, C., & Calantone, R. (2003). The effects of an integrative supply chain strategy on customer service and financial performance: an analysis of direct versus indirect relationships. *Journal of Operations Management*, *21*(5), 523–539.

Wahlster, W., Helbig, J., Hellinger, A., Stumpf, M. A. V., Blasco, J., Galloway, H., & Gestaltung, H. (2013). *Recommendations for implementing the strategic initiative INDUSTRIE 4.0*. Retrieved from http://www.acatech.de/fileadmin/user_upload/Baumstruktur_nach_Website/Acatech/root/de/Material_fuer_Sonderseiten/Industrie_4.0/Final_report__Industrie_4.0_accessible.pdf

Wan, J., Cai, H., & Zhou, K. (2015). Industrie 4.0: Enabling technologies. *Proceedings of 2015 International Conference on Intelligent Computing and Internet of Things*, 135–140. https://doi.org/10.1109/ICAIOT.2015.7111555

Wan, J., Chen, M., Xia, F., Di, L., & Zhou, K. (2013). From machine-to-machine communications towards cyber-physical systems. *Computer Science and Information Systems*, *10*(3), 1105–1128.

Radanliev, Petar, David De Roure, Kevin Page, Jason R.C. Nurse, Rafael Mantilla Montalvo, Omar Santos, La'Treall Maddox, and Pete Burnap. "Cyber Risk at the Edge: Current and Future Trends on Cyber Risk Analytics and Artificial Intelligence in the Industrial Internet of Things and Industry 4.0 Supply Chains." *Cybersecurity, Springer Nature*, 2020. https://doi.org/10.1186/s42400-020-00052-8.

https://doi.org/10.2298/CSIS120326018W

Wang, L. (2013). Machine availability monitoring and machining process planning towards Cloud manufacturing. *CIRP Journal of Manufacturing Science and Technology*, 6(4), 263–273. https://doi.org/10.1016/j.cirpj.2013.07.001

Wang, L., Törngren, M., & Onori, M. (2015). Current status and advancement of cyber-physical systems in manufacturing. *Journal of Manufacturing Systems*, 37, 517–527. https://doi.org/10.1016/j.jmsy.2015.04.008

Wang, L., Wang, X. V., Gao, L., & Váncza, J. (2014). A cloud-based approach for WEEE remanufacturing. *CIRP Annals - Manufacturing Technology*, 63(1), 409–412. https://doi.org/10.1016/j.cirp.2014.03.114

Wang, S., Wan, J., Li, D., & Zhang, C. (2016). Implementing Smart Factory of Industrie 4.0: An Outlook. *International Journal of Distributed Sensor Networks*, 12(1), 1–10. https://doi.org/10.1155/2016/3159805

Wang, Y., Wu, W., Zhang, C., Xing, X., Gong, X., & Zou, W. (2019). From proof-of-concept to exploitable. *Cybersecurity*, 2(1), 1–25. https://doi.org/10.1186/s42400-019-0028-9

Wark, T., Corke, P., Sikka, P., Klingbeil, L., Guo, Y., Crossman, C., … Bishop-Hurley, G. (2007). Transforming Agriculture through Pervasive Wireless Sensor Networks. *IEEE Pervasive Computing*, 6(2), 50–57. https://doi.org/10.1109/MPRV.2007.47

Weyer, S., Schmitt, M., Ohmer, M., & Gorecky, D. (2015). Towards Industry 4.0 - Standardization as the crucial challenge for highly modular, multi-vendor production systems. *IFAC-PapersOnLine*, 48(3), 579–584. https://doi.org/10.1016/j.ifacol.2015.06.143

World Economic Forum. (2015). *Partnering for Cyber Resilience Towards the Quantification of Cyber Threats*. Retrieved from http://www3.weforum.org/docs/WEFUSA_QuantificationofCyberThreats_Report2015.pdf

Yen, B., Farhoomand, A., & Ng, P. (2004). Constructing an e-Supply Chain at Eastman Chemical Company. *Journal of Information Technology*, 19(2), 93–107. https://doi.org/10.1057/palgrave.jit.2000011

Zhang, Q., Jia, S., Chang, B., & Chen, B. (2018). Ensuring data confidentiality via plausibly deniable encryption and secure deletion – a survey. *Cybersecurity*, 1(1), 1–20. https://doi.org/10.1186/s42400-018-0005-8

Zhu, Q., Rieger, C., & Basar, T. (2011). A hierarchical security architecture for cyber-physical systems. *2011 4th International Symposium on Resilient Control Systems*, 15–20. https://doi.org/10.1109/ISRCS.2011.6016081

---

[i] https://www.petrashub.org